\documentstyle[12pt,a4]{article}

\def\<{\langle}
\def\>{\rangle}
\def\/{\sqrt{}}
\def\a{\alpha}
\def\b{\beta}
\def\d{\delta}
\def\l{\lambda}
\def\e{\epsilon}

\begin{document}

\vskip 1.5 cm 
\centerline{\Large 
Spectral Evolution of the Universe}
\vskip 1.0 cm 

\centerline{\large {\sc Masafumi Seriu}}
\vskip .2cm
\centerline{\it Institute of Cosmology}
\centerline{\it Department of Physics \& Astronomy}
\centerline{\it Tufts University}
\centerline{\it Massachusetts 02155, USA~\footnote{
Present address. E-mail: mseriu@cosmos2.phy.tufts.edu}}
\centerline{*} 
\centerline{\it  Department of Physics, Fukui University}
\centerline{\it  Fukui 910-8507, Japan~\footnote{
Permanent address. E-mail: mseriu@edu00.f-edu.fukui-u.ac.jp}}
\centerline{*} 
\centerline{\it Yukawa Institute for Theoretical Physics}
\centerline{\it Kyoto University}
\centerline{\it Kyoto 606-8224, Japan}
\vskip 1.5cm

\begin{abstract}
We derive the evolution equations for  the spectra of a space 
(``Universe"). Here the spectra mean the eigenvalues of the 
Laplacian on a space, which  contain the geometrical information 
on the space. 

 These spectral evolution equations are expected to be  useful to 
analyze the time evolution of the geometrical structures of the Universe. 
In particular, it is indispensable to investigate the time evolution of  
  the spectral distance between two spaces, which is a measure of closeness 
between two geometries defined in terms of the spectra.  

As an application, we investigate the time evolution of the spectral distance 
between two  Universes that are very close to each other; 
it is  the first necessary step for
 analyzing the time evolution of the geometrical discrepancies between 
the real Universe and its model.   We find out a universal formula for 
the spectral distance between two very close Universes, which turns out to be  
independent of the detailed form of the spectral distance nor 
the gravity theory. Then we investigate its time evolution with the help of 
 the spectral evolution equations. We also formulate the criteria 
 for a good cosmological model in terms of the spectral distance.
\end{abstract}

\section{Introduction}
\label{section:I}

The notion of {\it closeness} or {\it distance}
 plays an essential role in  
 physics.  We frequently encounter the  concept 
 in the course of  constructing and applying a theory.
 
 The first step of finding out a law 
 of nature is often  classifying the objects 
 in question into several categories (e.g., Hubble's classification  
 of galaxies). 
  For classification, we are tacitly assuming  the concept of 
  closeness between the objects to be classified.   
Next,  when we judge the validity of a theory by experiment, we 
construct a suitable parameter space associated to  the theory and check  
 the  closeness between  the  point predicted by the theory 
  and the experimental data-point in the parameter space. 
  Here some kind of distance in the parameter space should be assumed.  
Finally,  when we try to explain an observed phenomenon  
 by  a certain model based on   a more or less established  theory, 
 we need to compare the observational data 
 with the predictions of the model (e.g. the relation between 
 a signal of the gravitational wave and its templates; 
  a comparison of the observational data 
 with the results of numerical simulations based on a model).  
  Again some notion of distance is 
 required in a parameter space suitable for this purpose. 

 Here we find out  a universal setting for comparing 
 ``theory and reality":  
 There should be  a parameter space equipped with the  notion of closeness. 
 We prepare a set of models (or templates). 
 Each model corresponds to a point in the parameter space, so that 
 the points corresponding to these models  are distributed over 
 the parameter space. 
 Now, the observed data define  another point in the parameter space. 
 Then we try to find out the model-point that is closest to the data-point.  
 
 Thus we realize the significance of establishing 
 a parameter space and a distance/closeness on it according to 
 the problem we need to study.

The same situation occurs in cosmology and  spacetime physics. 
Indeed, we often need to consider a set of 
Universes, rather than  just only ``our Universe": 
Cosmology itself is a trial for grasping overall, averaged nature of 
the complicated reality of our Universe 
in terms of models; when we question why our Universe emerged 
rather than other possibilities, we are considering 
a set of Universes.    
According to the above considerations, thus, it is essential to 
establish a space of all Universes and a distance/closeness between 
any two Universes among them. 

In a series of investigations, it turned out that 
we can in fact  construct a space of all compact 
Universes equipped with a sort of distance.  

Let  $Riem$ be  the space
 of all $(D-1)$-dimensional, compact Riemannian geometries without 
 boundaries\footnote{
 Throughout this paper,  $D$  represents  
the spacetime dimension, so that the spatial section  is 
always $(D-1)$-dimensional.
 }. On $Riem$,  we can introduce a measure of closeness in terms of 
the spectra, a set of eigenvalues of a certain elliptic 
operator~\cite{MS-spectral}. 
Here we consider only the Laplacian  $\Delta$ as 
an elliptic operator. For a given geometry 
${\cal G} \in Riem $, we get the {\it spectra}, or 
a set of eigenvalues of the Laplacian 
$\{\l_n\}_{n=0}^\infty=$
$\{0=\l_0 < \l_1 \leq \l_2 \leq \cdots \leq \l_n \leq \cdots 
\uparrow \infty \}$, 
numbered in an increasing order. Since $\l_n$ has  dimension 
$[{\rm Length}^{-2}]$, 
the higher (lower) spectrum in general reflects the smaller (larger) 
scale properties of the geometry.
Therefore the  spectra are  desirable quantities for 
describing the effective geometrical structures of the space at each 
observational scale, 
e.g. the scale-dependent topology~\cite{Vis,MS-scale}.   
Let us call this type of  representation of geometry in terms of the spectra 
the {\it spectral representation} of geometrical 
structures~\cite{MS-spectral}.

Suppose $\cal G$ and ${\cal G}'$ are two spaces in $Riem$,  and 
let $\{\l_n\}_{n=0}^\infty$ and $\{\l'_n\}_{n=0}^\infty$ be 
the spectra for $\cal G$ and ${\cal G}'$, respectively. 
By comparing $\{\l_n\}_{n=1}^N$ and $\{\l'_n\}_{n=1}^N$ 
in a suitable manner, we can introduce a measure of closeness 
between $\cal G$ and ${\cal G}'$ of order $N$~\cite{MS-spectral}. 
It compares two geometries up to the scale of order $O(\l_N^{-1/2})$, 
neglecting the smaller scale differences (we shall discuss 
more in detail on this topic 
in  \ref{subsection:IV-2} and \ref{subsection:IV-3}).  

However, it  turns out that this measure of closeness 
$d_N({\cal G}, {\cal G}')$ does not satisfy the triangle inequality 
though it satisfies the other two axioms of 
distance~\cite{MS-spectral}.   
Even though the triangle inequality is far from  a {\it must}
 from the viewpoint 
of the general theory of point set topology, 
it is certain that the inequality makes several arguments concise and 
it makes the measure of closeness more compatible with our notion of closeness.

This problem has been  resolved by realizing that  the breakdown of the 
triangle inequality is a  mild one in a certain sense,  and proving  
that $Riem$ equipped with $d_N({\cal G}, {\cal G}')$ forms 
 a metrizable space~\cite{MS-space}. In other words, it has been justified to  
regard $d_N({\cal G}, {\cal G}')$  as a distance provided that a care is taken 
when the triangle inequality is required in the argument; we also found out  
 the alternative of $d_N({\cal G}, {\cal G}')$  to be used when 
the triangle inequality is needed.  
(See \ref{subsection:IV-2} and  Ref.\cite{MS-space} for more details.)
As an immediate consequence, we can even introduce a distance on 
$Slice({\cal M}, g)$, 
a space  of all possible time-slices of a given spacetime $({\cal M}, g)$. 
(See Ref.\cite{MS-AVE} for more details on this point and  
its application to the averaging problem in cosmology~\cite{AVE,AVE2}.) 
From now on, we call $d_N({\cal G}, {\cal G}')$ the 
{\it spectral distance}. 

Following  the arguments at the beginning of this section, 
we have established a parameter space and a distance on it
appropriate for spacetime physics:  
The space of all spaces of order $N$, ${\cal S}_N$, 
which is  a completion of $(Riem, d_N)/_\sim $ is what we had been 
searching for. (Here $/_\sim$ indicates the identification of 
isospectral manifolds~\cite{CH}. We discuss  a physical interpretation of 
the isospectral manifolds in Section \ref{section:V}. 
See also Ref.\cite{MS-space,MS-JGRG}.) 
Because of its nice property,  ${\cal S}_N$ can be regarded as 
 a basic arena for the study of spacetime physics. For instance, 
 we can define integral over ${\cal S}_N$~\cite{MS-space}, 
 which would be 
 essential in   quantum cosmology.

Being ${\cal S}_N$ at hand,  
we are now in a position to handle a set of Universes.  
For definiteness, let us focus on cosmological problems now. 
In cosmology, we need to judge to what extent a model reflects 
the real Universe. There is no guarantee whether cosmology 
is possible, viz. whether a model close to the reality at some instant of time, 
remains so all the time.   From the viewpoint of the 
spectral representation, this fundamental problem 
 can be visualized as follows: 
Let $\cal G$ be the real Universe at present with respect to 
 a certain time-slicing. 
 Let ${\cal G}'$ be a model located in the  
 neighborhood of $\cal G$ in  ${\cal S}_N~$\footnote{Here a model means  
 a spacetime (usually it possesses much simpler geometrical structures 
 than reality)  along with a certain fixed time-slicing. We here regard 
  an identical spacetime with different time-slicings as 
 two different models.}.    Then we should investigate 
  the time evolution of $d_N({\cal G}, {\cal G}')$ and   
 should  analyze in what conditions $d_N({\cal G}, {\cal G}')$ 
 remains small during a certain period of time. 

This kind of investigation is now possible since the spectral distance is 
defined explicitly in terms of the spectra, 
which have a firm basis both physically and 
mathematically. What we now  need is, thus, to analyze 
 the time evolution of  the spectra. 
In the spectral representation, the spectra $\{\l_n \}$ are 
placed in the most fundamental position. Thus, from the purely theoretical 
viewpoint, too, it is interesting to investigate 
 the time evolution of the spectra in detail. 

As a first step, we can understand how the time evolution of the spectra 
is induced by the evolution of geometry  as follows: 
By evolving an initial $(D-1)$-dimensional geometry $(\Sigma , h)$ 
 according to the Einstein equations (in the Hamiltonian form if necessary),  
we get a 1-parameter family of geometries $(\Sigma , h(t))$.
In principle, then, we can get the spectra for each geometry $(\Sigma , h(t))$. 
In this manner, we get a 1-parameter family of sets of spectra  $\{\l_n(t)\}$. 

It is more preferable both theoretically and practically, however, 
if  the time evolution of 
$\{\l_n(t)\}$ is described (1) solely in terms of spectral quantities, without 
any explicit reference to the metrical information behind them, and 
(2) in the form of differential equations of $\{\l_n(t)\}$ 
with respect to time.      

 The key procedure for  achieving this goal is to  
investigate the response of $\{\l_n \}$ to the change of the spatial 
metric $h$. Since the latter is controlled by the Einstein equations, 
we thus  expect to obtain the spectral version of the Einstein equations. 
The main aim of this paper is  to 
obtain the fundamental evolution equations for the spectra, by 
putting  this program into practice.

In section \ref{section:II}, we prepare several formulas that are 
needed in the subsequent investigations. In section  \ref{section:III}, 
which is the main part of this paper, 
we derive the spectral evolution equations. 
As basic applications of the results we obtained, we discuss three topics in 
section \ref{section:IV}. 
In \ref{subsection:IV-1}, we study the spectral evolution of the 
closed Friedmann-Robertson-Walker Universe. In \ref{subsection:IV-2}, 
we study the spectral distance between two Universes
 that are very close to each 
other in ${\cal S}_N$. We find out its universal expression in the leading 
order, which is independent of the detailed form of 
the spectral distance nor the gravity theory.   In \ref{subsection:IV-3}, 
we investigate the time evolution of the spectral distance between two very 
close Universes. 
Section \ref{section:V} is devoted for discussions.

\section{Basic formulas for the spectra}
\label{section:II}

Let $(\Sigma, g)$ be a $(D-1)$-dimensional compact  Riemannian manifold 
without boundaries. 
 We set an eigenvalue problem for the Laplacian  $\Delta$ on $({\cal M}, g)$,  
$\Delta f = -\l f$. Let $\{ \l_n \}_{n=0,1,2,\cdots}$
$:= \{ 0=\l_0 < \l_1 \leq \l_2 \leq \cdots \leq \l_n$$ 
\leq \cdots \uparrow  \infty \}$ be  the set of eigenvalues, or {\it spectra} 
hereafter, arranged in an increasing order. 
For simplicity of formulas, we assume that there is no degeneracy 
in the spectra throughout this paper.    
Let $\{ f_n  \}_{n=0,1,2,\cdots}$ be the set of 
real-valued eigenfunctions 
that are normalized as
\begin{equation} 
(f_m,\ f_n):=\int_\Sigma\ f_m \ f_n \  \/ = \d_{mn}\ \ , 
\label{eq:normalization}
\end{equation}
where the natural integral measure on $(\Sigma, g)$ 
is implied  by $\sqrt{}:=\sqrt{\det (g_{ab})}$.   
Let us note that a set 
$\{\frac{1}{\sqrt{\l_n}} \partial_a f_n \}_{n=1}^\infty$  
forms an orthonormal subset of 1-forms on $(\Sigma, g)$, 
\begin{equation} 
\frac{1}{\sqrt{\l_m \l_n}}
\int_\Sigma\  
 \partial_a f_m \  g^{ab}\   
  \partial_b f_n \  \/ = \d_{mn}\ \ . 
\label{eq:normalization-1form}
\end{equation}

\subsection{Spectral components of functions and tensors}
Let   $A$ and $A_{ab}$ be  any function and any 
symmetric tensor field, respectively,  on $(\Sigma , g)$.     
It is useful to introduce diffeomorphism invariant quantities 
$\< A {\>_{}}_{mn}$ and $\< A_{ab} {\>_{}}_{mn}$ defined  as
\begin{eqnarray*}
\< A {\>_{}}_{mn}
       &:=& \int_{\Sigma}\ f_m \  A \ f_n \ \/ \ \ , \\
\< A_{ab} {\>_{}}_{mn}
       &:=& \frac{1}{\sqrt{\l_m \l_n}}\int_{\Sigma} \ 
                \ \partial^af_m \  A_{ab} \   \partial^b f_n \ \/  \ \ .      
\end{eqnarray*}
Here, for a quantity of type $\< A_{ab}{\>_{}}_{mn}$, we always understand 
that  $n, m \geq 1$ unless otherwise stated.   
We note that Eqs.(\ref{eq:normalization}) and 
(\ref{eq:normalization-1form}) can be expressed as 
\begin{equation}
\<1{\>_{}}_{mn}=\d_{mn}\ , \ \ \<g_{ab} {\>_{}}_{mn}=\d_{mn} \ \ .  
\label{eq:identity}
\end{equation}
We also employ an abbreviated  notation 
\[ 
\< A {\>_{}}_{n}:=\< A {\>_{}}_{nn} \ , 
      \ \< A_{ab} {\>_{}}_{n}:=\< A_{ab} {\>_{}}_{nn}\ \ .
\]

For later uses, we develop these notations to 
\begin{eqnarray*} 
\< A {\>_{}}_{lmn}:&=&\< A f_l {\>_{}}_{mn}\ \ , \ \  
\< A {\>_{}}_{klmn}:=\< A f_k {\>_{}}_{lmn}\ , \cdots \ \ ,    \\
\< A_{ab} {\>_{}}_{l,mn}:&=&\< A_{ab} f_l {\>_{}}_{mn}\ \ , \ \ 
\< A_{ab} {\>_{}}_{kl,mn}:=\< A_{ab} f_k {\>_{}}_{l,mn}\ , \cdots \ \ . 
\end{eqnarray*}

We note that, for arbitrary functions $A$ and $B$ 
\begin{equation}
\<AB{\>_{}}_{mn}= \sum_{k=0}^\infty \<A{\>_{}}_{mk}\<B{\>_{}}_{kn} \ \ . 
\label{eq:<AB>}             
\end{equation}
To show this formula,  we   insert  the $\d$-function 
 into the integral 
expression of $\<AB{\>_{}}_{mn}$,  noting that  
$\d (x,y)\/_y^{-1}  = \sum _{k=0}^\infty f_k(x)f_k(y) $.  

The following formula  is also useful, which relates 
$\<A g_{ab}{\>_{}}_{mn}$  with $\<A{\>_{}}_{mn}$:
\begin{equation}
\<A g_{ab}{\>_{}}_{mn} 
      = \frac{\l_m + \l_n}{2\sqrt{\l_m \l_n}} \<A {\>_{}}_{mn}
          +   \frac{1}{2\sqrt{\l_m \l_n}} \<\Delta A {\>_{}}_{mn}\ \ .       
\label{eq:<Ag>}             
\end{equation}
To show this formula, one modifies    
the defining equation for $\<A g_{ab}{\>_{}}_{mn}$ with the help of 
 partial integrals, noting that the R.H.S. (right-hand side) should be 
 symmetric in $m$ and $n$ .
  
Setting $m=n$ in Eq.(\ref{eq:<Ag>}), we get  
\begin{equation}
\<A g_{ab}{\>_{}}_{n} 
      = \<A {\>_{}}_{n}
            +   \frac{1}{2 \l_n} \<\Delta A {\>_{}}_{n}\ \ .          
\label{eq:<Ag>-2}             
\end{equation}

\subsection{The variation formulas} 
We frequently  consider the variation $\d Q$ of a certain quantity 
$Q$ below. 
Here we  treat $\d$ as  a general variation for the time being. 
In later applications, 
the time-derivative (the Lie derivative along a time-flow 
 vector $t^a$ in a spacetime 
picture) is mostly considered as the variation operator $\d$. 

Now, noting that $\Delta f= \frac{1}{\/}\partial_a(\/\ g^{ab}\partial_b f)$ 
for  an arbitrary function $f$,  
 the variation of $\Delta$ is represented as\footnote{
  To avoid  trivial indices, 
we flexibly adopt  notations such as $\vec{u}:=u^a$, 
$A \cdot B := A_{ab}B^{ab}$, $\vec{u}\cdot \vec{v}:=u_a\ v^a$ and 
$A_\cdot^{\ \cdot}:=A_a^{\ a}$ .  
We also flexibly choose   symbols for the 
derivative of a function $f$, such as $D_a f=\vec{D}f =\partial_a f$.
(Here $D_a$ and $\vec{D}$ denote the covariant derivative.)
 } 
\[
\d \Delta f = \frac{1}{2}\partial_a (g\cdot \d g)\partial^a f 
-\frac{1}{\/} \partial_a (g^{ab} \d g_{bc}\partial^c f\/) \ \ . 
\]

Thus, employing the same kind of notation as $\<A{\>_{}}_{mn}$,  
we can introduce the quantity 
\[
\<\d \Delta{\>_{}}_{mn}:= \int\  f_m\  \d \Delta\  f_n\  \/ 
       = \frac{1}{2}\int f_m \partial_a (g\cdot \d g) \partial^a f_n \/ 
            - \int f_m \partial_a (g^{ab} \d g_{bc}\partial^c f_n \/)\ \ , 
\]
where we  note that 
the variation is taken only for the operator $\Delta$, and not 
for the eigenfunctions $f_n$, $f_m$. We should also keep in mind that 
 $\<\d \Delta {\>_{}}_{mn}$ is {\it not} symmetric in $m$ and $n$, unlike 
 $\<A {\>_{}}_{mn}$, because $\Delta$ is an operator,  and not a function.  
Noting that $f_0$ is a constant function 
(see {\it Appendix} \ref{section:Appendix A}), 
it is evident that $\<\d \Delta {\>_{}}_{m0}=0$ and 
$\<\d \Delta {\>_{}}_{0m}=\frac{\l_m}{2}\<g\cdot \d g{\>_{}}_{0m}$ 
 ($m=0,1,2, \cdots$).

With these preliminaries, we now investigate the variations of 
spectral quantities. We start with the variation of the spectra. 
From Eq.(\ref{eq:B-10}) in  {\it Appendix} \ref{section:Appendix B},  
it follows that 
\begin{equation}
\d \l_n = - \<\d \Delta {\>_{}}_{n}  \ \ , 
\label{eq:dlambda}             
\end{equation}
 which is a basic result of the perturbation theory 
 (``Fermi's golden rule'').

Now let us investigate  the variation of the eigenfunctions. 
We  have a general formula Eq.(\ref{eq:B-11}) with Eq.(\ref{eq:B-12}) (see 
{\it Appendix} \ref{section:Appendix B}) for the perturbation of 
eigenvectors. Here we need to specify the 
factor $c_n^{(1)}$ in Eq.(\ref{eq:B-12}).  For this purpose, 
we take the variation of 
the both sides of  Eq.(\ref{eq:normalization}) for $m=n$. Noting that 
$\d \/ = \frac{1}{2}g\cdot \d g \/$,   we get 
\begin{equation}
( f_n , \ \d f_n ) =-\frac{1}{4}\<g\cdot \d g {\>_{}}_{n}\ \ .
\label{eq:df-f}
\end{equation}
In the standard  perturbation analysis in  quantum mechanics, 
 the inner-product is fixed, and  not perturbed, while  
 the eigenfunctions are  perturbed. Thus, 
$\d f_n$ should be perpendicular to $f_n$ if $f_n$ is normalized.
In our case, on the other hand, the inner-product is also subject to 
the variation  because of the presence of  the integral measure $\/$.
 Thus $\d f_n$ is not in general perpendicular 
to $f_n$, as is clear in Eq.(\ref{eq:df-f}). 
Combining  Eq.(\ref{eq:df-f}) with 
Eqs.(\ref{eq:B-11}) and Eq.(\ref{eq:B-12}), 
we see that $c_n^{(1)}$ in Eq.(\ref{eq:B-12}) 
should be chosen as $c_n^{(1)}=-\frac{1}{4}\<g\cdot \d g {\>_{}}_{n}$ 
in the present case. Thus, we get 
\begin{equation}
\d f_n = \sum_{k=0}^\infty f_k\  \mu_{kn} \ \ ,
\label{eq:df}             
\end{equation}
where 
\begin{equation}
\mu_{mn}:=
        \left\{
        \begin{array}{ll}
        \frac{\< \d \Delta {\>_{}}_{mn}}{\l_m - \l_n} \ \ 
                             & \mbox{for}\ \ m \neq n  \nonumber \\
         -\frac{1}{4}\<g \cdot \d g \>_m   \ \ 
                             & \mbox{for}\ \ m = n\ \ . \\
       \end{array}
        \right.  
\label{eq:mu}
\end{equation}
Taking the inner-product of the both sides of Eq.(\ref{eq:df}) with $f_m$, 
we get 
\begin{equation}
   \mu_{mn}= (f_m, \  \d f_n) \ \ ,
\label{eq:fdf}
\end{equation}
which gives a clear interpretation of the quantity $\mu_{mn}$ as 
the projection of $\d f_n$ to the direction of $f_m$.

With the help of Eq.(\ref{eq:df}), it is now straightforward to show that 
\begin{equation} 
\d \< A {\>_{}}_{mn} = \<\d A{\>_{}}_{mn} 
                        + \frac{1}{2}\<A\  g\cdot \d g{\>_{}}_{mn} 
                            +\sum_k \< A{\>_{}}_{mk}\ \mu_{kn }
                            +\sum_k \< A{\>_{}}_{nk}\ \mu_{km }\ \ . 
\label{eq:d<A>}
\end{equation}
In particular, for the case of $m=n$, we get 
\begin{equation} 
\d \< A {\>_{}}_{n} = \<\d A{\>_{}}_{n} 
                       + \frac{1}{2}\<A\  g\cdot \d g{\>_{}}_{n} 
                            +2 \sum_k \< A{\>_{}}_{nk}\ \mu_{kn } \ \ . 
\label{eq:d<A>-2}
\end{equation}

Introducing 
$\Gamma_{mn}:= -\mu_{mn} - \frac{1}{4}\<g \cdot \d g {\>_{}}_{mn} $ 
(note that $\Gamma_{nn}=0 $),  
Eq.(\ref{eq:d<A>}) is also represented as 
\[
\d \< A {\>_{}}_{mn} 
     = \<\d A{\>_{}}_{mn}  
        -\sum_k \< A{\>_{}}_{mk} \Gamma_{kn}
                            -\sum_k \< A{\>_{}}_{nk} \Gamma_{km }\ \ .
\]
It is interesting that this expression is in  a similar form as 
the covariant derivative of a symmetric tensor. 

In the same manner, we get a formula for $\d \<A_{ab}{\>_{}}_{mn}$ as 
\begin{eqnarray} 
\d \< A_{ab} {\>_{}}_{mn} 
        &=& \<\d A_{ab}{\>_{}}_{mn} 
                  -(\d \ln \sqrt{\l_m \l_n}\ ) \< A_{ab} {\>_{}}_{mn}     
                  + \frac{1}{2}\<A_{ab}\  g\cdot \d g{\>_{}}_{mn} 
                 -\< A_a^{\ c}\d g_{cb} + A_b^{\ c}\d g_{ca} {\>_{}}_{mn} 
                                         \nonumber  \\
        &&\ \ \  +\sum_k \sqrt{\frac{\l_k}{\l_n}}\< A_{ab}{\>_{}}_{mk}\ \mu_{kn}
           +\sum_k \sqrt{\frac{\l_k}{\l_m}}\< A_{ab}{\>_{}}_{nk}\ \mu_{km}\ \ , 
\label{eq:d<Aab>}
\end{eqnarray}
and for the case of $m=n$, we get 
\begin{eqnarray} 
\d \< A_{ab} {\>_{}}_{n} 
        &=& \<\d A_{ab}{\>_{}}_{n} 
                  -(\d \ln  \l_n ) \< A_{ab} {\>_{}}_{n}     
                  + \frac{1}{2}\<A_{ab}\  g\cdot \d g{\>_{}}_{n} 
                  -2\< (A\d g)_{ab}{\>_{}}_{n}         \nonumber  \\
     &&\ \ \  + 2\sum_k \sqrt{\frac{\l_k}{\l_n}}\< A_{ab}{\>_{}}_{nk}\ \mu_{kn} 
                                                         \ \ .  
\label{eq:d<Aab>-2}
\end{eqnarray}

\subsection{Basic identities}
\label{subsection:II-3}
 We obtain important relations by taking the variation of the basic 
 identities Eq.(\ref{eq:identity}). 
 
 First, we take  the variation of the both sides of $\d_{mn}=\<1{\>_{}}_{mn}$.
 Then, with the help of Eq.(\ref{eq:d<A>}), we get  
\begin{eqnarray*}
0= \frac{1}{2}\<g\cdot \d g {\>_{}}_{mn} +  \mu_{mn} +  \mu_{nm}   \ \ ,     
\end{eqnarray*} 
which implies  an identity, 
\begin{equation}
\<g\cdot \d g {\>_{}}_{mn} = -2 (\mu_{mn} +  \mu_{nm}) \ \ .
\label{eq:<gdg>1}
\end{equation} 

Thus, with the help of Eq.(\ref{eq:mu}), we get
\begin{equation}
\<g\cdot \d g {\>_{}}_{mn}
         = -\frac{2}{\l_m - \l_n}
          (\<\d \Delta {\>_{}}_{mn}-\<\d \Delta {\>_{}}_{nm})\ \ {\rm for}\ \ 
                                  m \neq n \ \ .
\label{eq:<gdg>2}
\end{equation} 
On the other hand, no condition is imposed on $\<g\cdot \d g {\>_{}}_{n}$ 
except for 
\begin{equation}
\<g\cdot \d g\>_{{}_{0}}= 2 \frac{\d V}{V}\ \ ,
\label{eq:<gdg>0} 
\end{equation}
which follows by an independent  argument 
(see {\it Appendix} \ref{section:Appendix A}).

Now we take the variation of the both sides of $\d_{mn}=\<g_{ab}{\>_{}}_{mn}$ 
for $m,n \geq 1$. Then, with the help of Eq.(\ref{eq:d<Aab>}), we get
\begin{equation}
0= - \<\overline{\d g}_{ab}{\>_{}}_{mn} - (\d \ln \l_n)  \d_{mn}
    + \sqrt{\frac{\l_m}{\l_n}}\ \mu_{mn}
    + \sqrt{\frac{\l_n}{\l_m}}\ \mu_{nm} \ \ ,
\end{equation} 
where $\overline{A}_{ab}:= A_{ab}-\frac{1}{2}A_\cdot^{\ \cdot} g_{ab}$\ \ for 
any symmetric tensor $A_{ab}$. 

Thus, we get
\begin{equation}
\<\overline{\d g}_{ab}{\>_{}}_{mn}=
    \left\{ 
    \begin{array}{ll}
           \sqrt{\frac{\l_m}{\l_n}}\ \mu_{mn} 
                       + \sqrt{\frac{\l_n}{\l_m}}\ \mu_{nm}
                        & {}                  \\ 
           \qquad  
                = \frac{1}{\l_m - \l_n} 
                  \left( \sqrt{\frac{\l_m}{\l_n}}\  \<\d \Delta {\>_{}}_{mn} 
                   - \sqrt{\frac{\l_n}{\l_m}}\  \<\d \Delta {\>_{}}_{nm}  
                  \right) 
                  &  \ \ {\rm for} \ \ m \neq n   \\
         - \d \ln \l_n \ - \frac{1}{2}\<g\cdot \d g \>_n \ \ 
                 &  \ \     {\rm for} \ \ m = n \ \ . \\                 
    \end{array}
    \right.     
\label{eq:<dg>-bar}
\end{equation}
The fact  that $\<\d \Delta {\>_{}}_{m0}=0$ ($m=0,1,2,\cdots$)  
suggests us to   formally define that 
$\<\overline{\d g}_{ab}{\>_{}}_{m0}=0$ ($m=0,1,2,\cdots$). We adopt this
formal definition here for a notational neatness (see Eq.(\ref{eq:<dDelta>})  
below).

From Eqs.(\ref{eq:<gdg>2}) and (\ref{eq:<dg>-bar}), we obtain 
a formula for $\< \d \Delta {\>_{}}_{mn}$, 
\begin{equation}
\< \d \Delta {\>_{}}_{mn}
         = \frac{\l_n}{2}\<g\cdot \d g {\>_{}}_{mn} 
               + \sqrt{\l_m \l_n}\<\overline{\d g}_{ab}{\>_{}}_{mn}\ \ .
\label{eq:<dDelta>}
\end{equation}
We can also derive   Eq.(\ref{eq:<dDelta>}) directly from 
the definition of $\< \d \Delta {\>_{}}_{mn}$. 

We note that Eq.(\ref{eq:<dDelta>}) is valid  for the case of $m=n$ also, 
due to Eq.(\ref{eq:dlambda}) and 
the second equation of Eq.(\ref{eq:<dg>-bar}). Furthermore, we  
realize that this formula is also valid  for $m=0$ or $n=0$ 
on account of the formal definition  
$\<\overline{\d g}_{ab}{\>_{}}_{m0}=0$ 
($m=0,1,2,\cdots$).

We now investigate a different type of identities. 

We pay attention to the second equation in Eq.(\ref{eq:<dg>-bar}), 
\begin{equation}
\d \ln \l_n = -\<\overline{\d g}_{ab}{\>_{}}_{n} 
                -\frac{1}{2}\<g\cdot \d g {\>_{}}_{n} \ \ . 
\label{eq:dloglambda}
\end{equation}
From this relation, it is straightforward to show a formula
\begin{equation}
\int A_{ab} \frac{\d \ln \l_n}{\d g_{ab}} 
   = -\<\overline{A}_{ab}{\>_{}}_{n} 
             - \frac{1}{2}\<A_\cdot^{\ \cdot} {\>_{}}_{n}\ \ ,
\label{eq:identityAab}
\end{equation}
where $A_{ab}$ is any symmetric tensor field. 

Let $u^a$ be any vector field.  
Substituting
 $A_{ab}={\cal L}_{\vec{u}}\ g_{ab}$, the L.H.S. (left-hand side) of 
Eq.(\ref{eq:identityAab}) vanishes because of the diffeomorphism invariance 
of the spectra, so that
\begin{eqnarray}
\<{\cal L}_{\vec{u}}\ g_{ab} {\>_{}}_{n}
   &=& \< \vec{D}\cdot \vec{u}\ g_{ab}\ {\>_{}}_{n} 
                  -\< \vec{D}\cdot \vec{u}{\>_{}}_{n} \ \ , \nonumber \\
\<\overline{D_{(a} u_{b)}}{\>_{}}_{n} 
           &+&   \frac{1}{2}\< \vec{D}\cdot \vec{u} {\>_{}}_{n}=0 \ \ .
\label{eq:identityDU}
\end{eqnarray}

We now note another identity.
Using the basic properties of the covariant derivative $D_a$, 
it is easily shown that 
\[
 \Delta D_a f = D_a \Delta f + \mbox{\boldmath $R$}_{ab}D^b f \ \ ,
\]
where $f$ is any smooth function. In particular, choosing 
$f_n$ as $f$, we get 
\begin{equation}
(g_{ab} \Delta  - \mbox{\boldmath $R$}_{ab})\partial^b f_n 
= -\l_n \partial_a f_n \ \ , 
\label{eq:delf}
\end{equation}
 i.e. $\partial_a f_n$ turns out to be an eigenfunction of 
 $g_{ab} \Delta  - \mbox{\boldmath $R$}_{ab}$ with the eigenvalue 
 $\l_n$. Taking the inner-product of  Eq.(\ref{eq:delf}) with $\partial_a f_m$, 
 we get 
\begin{equation}
\int D_aD_bf_m \  D^aD^b f_n \/ 
     = -\sqrt{\l_m \l_n} \<\mbox{\boldmath $R$}_{ab} {\>_{}}_{mn} 
                  + \l_n^2 \d_{mn}\ \ ,
\label{eq:<R>}                  
\end{equation}
or 
\begin{equation}
\< \mbox{\boldmath $R$}_{ab}{\>_{}}_{mn} 
      = \<  g_{ab} \Delta  {\>_{}}_{mn} + \l_n \d_{mn}\ \ . 
\label{eq:<R>2}                  
\end{equation}

\section{Evolution equations for the spectral quantities}
\label{section:III}

Now we investigate the time evolution of the spectra of a space. 
We have  prepared in the previous section  
 basic formulas regarding  
the responses of the spectral quantities 
 with respect to a change $\d g_{ab}$.
When we let $\d g_{ab}$ be of a dynamical origin, thus, we 
automatically obtain the evolution equations for the spectral quantities. 

We consider $(\Sigma, h)$, a $(D-1)$-dimensional compact  Riemannian manifold 
without boundaries, as a mathematical model of the spatial section 
of the Universe. For the present purpose,  we interprete a quantity 
$\d Q$ as $\frac{d}{d \a}_{|_\a=0} Q(\a)$, and identify the latter 
quantity  with 
$\dot{Q}:={\cal L}_{\vec{t}}\ Q $, where $\vec t$ is 
a time-flow vector\footnote{ 
See the argument at the end of {\it Appendix} {\ref{section:Appendix B}} 
also. }.
In particular, we replace   $\d g_{ab}$, $g \cdot \d g$ and 
$\overline{\d g}_{ab}$ in the 
previous section with corresponding  quantities as
\begin{eqnarray}
 \d g_{ab} &\longmapsto& \dot{h}_{ab}= 2 N K_{ab}+ 2 D_{(a}N_{b)}\ \ ,
                                         \nonumber  \\
 g \cdot \d g &\longmapsto& h^{ab}\dot{h}_{ab}
                      = 2NK + 2 \vec{D}\cdot \vec{N}\ \ ,
                                         \nonumber   \\
 \overline{\d g}_{ab} &\longmapsto& \overline{\dot{h}}_{ab}
         = 2N\overline{K}_{ab} + 2 \overline{D_{(a}N_{b)}}\ \ .
\label{eq:mapsto}         
\end{eqnarray}
Here  $K_{ab}$ is the extrinsic curvature and $K:=K_\cdot^{\ \cdot}$; 
 $N$ and $N_a$ are the lapse function and the shift vector, respectively.

\subsection{Evolution equation for $\{ \l_n \}$}

 First, Eq.(\ref{eq:dloglambda}) becomes 
\begin{eqnarray}
\dot{\l}_n &=& -\left( 
                      2 \<N \overline{K}_{ab}{\>_{}}_{n} 
                      +\<NK {\>_{}}_{n}
                \right) \l_n        
                                              \nonumber  \\
           &=& -2 \<N K_{ab}{\>_{}}_{n}\l_n 
                 +\frac{1}{2} \<\Delta (NK) {\>_{}}_{n} \ \ ,
\label{eq:lambda-dot}                        
\end{eqnarray}
where we used the identity Eq.(\ref{eq:identityDU}) in the first line, 
and Eq.(\ref{eq:<Ag>-2}) in the second line. We note that 
the shift vector $N_a$ does not appear in Eq.(\ref{eq:lambda-dot}), reflecting 
the spatial diffeomorphism invariance of $\l_n$.

Now, Eqs.(\ref{eq:<gdg>1}) and (\ref{eq:<dg>-bar}) becomes     
\begin{eqnarray*}
&& \<NK {\>_{}}_{mn} + \<\vec{D}\cdot \vec{N} {\>_{}}_{mn} 
       = -(\mu_{mn}+\mu_{nm})\ \ \nonumber \\
&& \<N \overline{K}_{ab}{\>_{}}_{mn} + \frac{1}{2}(\ln \l_n \dot{)}\ \d_{mn} 
        + \< \overline{D_{(a}N_{b)}} {\>_{}}_{mn} 
         = \frac{1}{2}
                 \left(\sqrt{\frac{\l_m}{\l_n}}\ \mu_{mn} 
                       + \sqrt{\frac{\l_n}{\l_m}}\ \mu_{nm}\right) \ , 
\end{eqnarray*}
where $\mu_{mn}$ is given by Eq.(\ref{eq:mu})  with $\< \d \Delta {\>_{}}_{mn}$
 replaced by $\< \dot{\Delta} {\>_{}}_{mn}$.

 Eq.(\ref{eq:<dDelta>}) becomes
\begin{equation}
\< \dot{\Delta} {\>_{}}_{mn}
= \l_n \<NK {\>_{}}_{mn} 
       + 2\sqrt{\l_m\l_n}\<N \overline{K}_{ab}{\>_{}}_{mn}
       +\l_n \<\vec{D}\cdot \vec{N} {\>_{}}_{mn}
       +2\sqrt{\l_m\l_n}\< \overline{D_{(a}N_{b)}}{\>_{}}_{mn}  \ \ .
\label{eq:<Delta-dot>}                        
\end{equation}

Ultimately we are only interested in the dynamics of the spectra 
$\{ \l_n \}$, which is invariant under the spatial diffeomorphism. 
The coordinate dependence of  other subsidiary 
variables like $\mu_{mn}$, which can dependent on the shift vector $N_a$, 
 should not have any influence on  the dynamics of $\{ \l_n \}$. 
Hence we can set $N_a=0$ from the very beginning to make formulas simpler.
 For simplicity we also set $N=1$ hereafter.

Now let us investigate  the evolution equations for the spectra, 
Eq.(\ref{eq:lambda-dot}),  in detail. 
At this stage, it is useful  to  develop  notations.    
First, we note that we can 
expand any scalar function $A(\cdot)$ in terms of $\{ f_n \}_{n=0}^\infty$: 
\begin{equation}
A(\cdot)=\sum_{l=0}^\infty A_n\ f_n(\cdot) \ \ .
\label{eq:fourier}                        
\end{equation}
Then, it follows that 
\[
\Delta A(\cdot)= -\sum_{l=1}^\infty \l_n A_n\ f_n(\cdot) \ \ ,  
\]
where we  see that the homogeneous component of $A$, $A_0$,
does not appear in the summation. 

The homogeneous component $A_0$ is related to the 
spatial average of $A$ over the spatial section $\Sigma$, 
$A_{\rm av}:=\frac{1}{V}\int_{{}_\Sigma} A \/ $, as 
\begin{equation}
 A_{\rm av}=A_0/\sqrt{V}\ \ . 
\label{eq:average}                        
\end{equation}

Next, let us introduce the quantity 
\begin{equation}
(l\ m\ n):= \<f_m {\>_{}}_{ln}=\<1 {\>_{}}_{mln}=\int\ f_l\ f_m\ f_n \/ \ \ .
\label{eq:(lmn)}                        
\end{equation}
Note that $(l\ m\ n)$ is totally symmetric in $l$, $m$  and $n$. 
We can also introduce a similar quantity 
 $(l\ m\ n\ k):=\int\   f_l\  f_m\  f_n\   f_{k}\  \/$ , and similarly  
 $(l\ m\ n\ k\ h)$,  and so on. However, the quantities of the type 
 $(l\ m\ n)$ are sufficient since 
  other quantities  can be  expressed in terms of $(l\ m\ n)$. For instance,   
\[
(l\ m\ n\ k)=\frac{1}{3}\sum_{l'}
 \left\{(l\ m\ l')(l'\ n\ k) + (m\ n\ l')(l'\ k\ l) 
 + (n\ k\ l')(l'\ l\ m) \right\}\ \ ,
\]
due to Eq.(\ref{eq:<AB>}).

We  note that the quantity  of the form  $\<A {\>_{}}_{lmn\cdots}$ 
is represented in terms of $A_l$ and $(l\ m\ n)$ since 
\[
\<A {\>_{}}_{lmn\cdots} = \sum_{l'} A_{l'} (l'\ l\ m\ n\ \cdots)\ \ . 
\]

In the same manner, the quantities of the form
 $\<A_{ab}{\>_{}}_{kl\cdots,mn}$ is represented in terms of  
$\<A_{ab} {\>_{}}_{l,mn}$ and $(l\ m\ n)$ because of the relation    
\[
\<A_{ab} {\>_{}}_{kl\cdots,mn}=\sum_{l'}
\<A_{ab} {\>_{}}_{l',mn}(l'\ k\ l\ \cdots)\ \ . 
\]
In particular, the quantity of the form  $\<A_{ab} {\>_{}}_{mn}$ 
can be represented in terms of  $\<A_{ab} {\>_{}}_{0,mn}$ as    
\[
\<A_{ab} {\>_{}}_{mn} = \<A_{ab} {\>_{}}_{0,mn}\sqrt{V}\ \ .
\]

Keeping  the applications to cosmology in mind, 
it is also useful to introduce the quantities $\e_{ab}$ and $r_{ab}$
 defined as 
\begin{eqnarray}
\e_{ab} &:=& K_{ab}-\frac{1}{D-1}K h_{ab} \ \ ,\ \ \nonumber \\
r_{ab} &:=&
  \mbox{\boldmath $R$}_{ab}- \frac{1}{D-1}\mbox{\boldmath $R$}h_{ab} \ \ .
\label{eq:e/r}                        
\end{eqnarray}
Note that $\e_\cdot^{\ \cdot} = r_\cdot^{\ \cdot} =0$. The 
quantities $\e_{ab}$ and $r_{ab}$ characterize the deviation  
of the spatial geometry $(\Sigma, h)$ from the isotropic geometry.

Now,  we note that 
\[
\<K{\>_{}}_{mn}=\sum_l K_l \  (l\ m\ n)\ \ ,
\]
and 
\begin{eqnarray*}
\< \overline{K}_{ab}{\>_{}}_{mn}
 &=& -\frac{D-3}{2(D-1)} \< K h_{ab}{\>_{}}_{mn} + \< \e_{ab}{\>_{}}_{mn} \\
 &=& -\frac{D-3}{4(D-1)} \frac{1}{\sqrt{\l_m \l_n}} 
                   \sum_l (\l_m + \l_n - \l_l)\ K_l\  (l\ m\ n) 
                             + \< \e_{ab}{\>_{}}_{mn}\ \ . 
\end{eqnarray*}
Here Eq.(\ref{eq:fourier}) has been applied to $K$ and 
 we have used Eq.(\ref{eq:<Ag>}).
  
Thus,  Eq.(\ref{eq:lambda-dot}) is  represented as    
\begin{equation}
\dot{\l}_n
= -\frac{2}{D-1} \sum_{l} (\l_n + \frac{D-3}{4}\l_l)\ K_l \  (l\ n\ n) 
              - 2\l_n \<\e_{ab} {\>_{}}_{n}  \ \ .
\label{eq:dlambda-2}                        
\end{equation}
This equation is also valid for $n=0$, viz. it is 
 compatible with $\l_0 \equiv 0$.

Looking at the R.H.S. of Eq.(\ref{eq:dlambda-2}), 
it turns out  that 
we further need the equations for $(l\ m\ n \dot{)\ }$, $\dot{K}_l$ and 
$\<\e_{ab} \dot{{\>_{}}_{n}}$ \ . We investigate  them one by one below.

\subsection{Evolution equation for $(l\ m\ n)$}

Applying Eq.(\ref{eq:d<A>}) along with Eq.(\ref{eq:df}) to 
$\<f_l {\>_{}}_{mn}=(l\ m\ n)$, we get   
\begin{equation}
(l\ m\ n \dot{)\ }=\sum_{l'} 
   \left\{ (l\ m\  l')\ \mu_{l'n} + (m\ n\ l')\ \mu_{l'l} +  
    (n\ l\ l')\ \mu_{l'm} \right\} 
    + \sum_{l'} (l\ m\ n\ l')K_{l'}\ \ . 
\label{eq:d(lmn)}                        
\end{equation}
Here,  from Eq.(\ref{eq:mu}) with Eq.(\ref{eq:<Delta-dot>}) ($N=1$, $N_a=0$), 
$\mu_{mn}$ is given by   
\begin{equation}
\mu_{mn}=
      \left\{
      \begin{array}{ll}
      \frac{D+1}{2(D-1)} \frac{1}{\l_m - \l_n} 
      \sum_l \left(\l_n - \frac{D-3}{D+1}(\l_m - \l_l)\right)K_l (l\ m\ n) 
         +\frac{2\sqrt{\l_m\l_n}}{\l_m - \l_n}\<\e_{ab} {\>_{}}_{mn}         
            & \mbox{for}\ \ m \neq n  \nonumber \\
             &                    \nonumber \\
             -\frac{1}{2}\sum_l K_l (l\ m\ m) \ \                
                             & \mbox{for}\ \ m = n\ \ . \\
       \end{array}
        \right.  
\label{eq:mu_mn}                        
\end{equation}

\subsection{Evolution equation for $K_l$}

Now, since $K_l=(K,\ f_l)$, we get  
$\dot{K}_l=(\dot{K}, \ f_l)+(K, \dot{f}_l) 
+ (K, \ f_l \  K)$, resulting in 
\begin{equation}
\dot{K}_l = (\dot{K})_l 
+ \sum_{l'} K_{l'} \mu_{l'l} + \sum_{l'l''} K_{l'}K_{l''} (l'\ l''\ l) \ \ ,
\label{eq:Kl-dot}                        
\end{equation}
where Eq.(\ref{eq:fdf}) has been applied. 
To get the detailed expression for  the first term $(\dot{K})_l$, we recall 
basic formulas of the canonical Einstein equations. We note  
\begin{equation}
\dot{K} =-\frac{1}{2\a}\frac{D-1}{D-2}p 
          -\frac{D-3}{2(D-2)}
          (\mbox{\boldmath $R$}+ K^2+ \frac{D-1}{D-3}K \cdot K)
                +\frac{D-1}{D-2} \Lambda \ \ , 
\label{eq:K-dot}                        
\end{equation}
 along with the constraints
\begin{eqnarray}
\mbox{\boldmath $R$}- K \cdot K + K^2 
                      - \frac{1}{\a} \rho - 2\Lambda &=& 0 \ \ ,  \nonumber \\ 
D^b K_{ab}-D_a K + \frac{1}{2\a} J_a &=& 0 \ \ ,
\label{eq:constraints}                        
\end{eqnarray}
 where $\a:=\frac{c^3}{16\pi G}$ and $\Lambda$ is the cosmological constant.
 Here we define the quantities $\rho$, $p$, $J_a$ and $S_{ab}$ 
 in connection with the above equations:   
 With the help of the normal unit vector $n^\a$ of 
  the spatial section, 
 the energy-momentum tensor of matter 
 $T^{\a \b}$ can be  decomposed into three components:   
 $T^{\a \b} = T_{{}_{\perp \perp}}^{\ \a \b}+T_{{}_{\| \perp}}^{\ \a \b}+
     T_{{}_{\| \|}}^{\ \a \b} $, where each term has the following form, 
\[
T_{{}_{\perp \perp}}^{\ \a \b}= \rho n^\a n^\b \ \ , \ \ 
T_{{}_{\| \perp}}^{\ \a \b} = J^\a n^\b + J^\b n^\a \ \ , \ \ 
T_{{}_{\| \|}}^{\ \a \b} = S^{\a \b}\ \ .
\]
(The suffix $\perp$ implies   ``perpendicular to the space'' and  
$\|$ implies ``along the space".)
Here $J^\a$ and $S^{\a\b}$ are spatial quantities and they are 
uniquely identified to their spatial counterparts, 
$J^a$ and $S^{ab}$, respectively.  
Then $\rho$, $J_a$ and  $S_{ab}$ are interpreted as the energy density, 
the momentum density and the stress tensor of matter, respectively, and 
$p:=\frac{1}{D-1}S_a^{\ a}$ defines the pressure of matter. 
(The vector and tensor indices 
of spatial quantities in this context are lowered and raised 
by, respectively,  the spatial metric $h_{ab}$ and its inverse $h^{ab}$.)

Now, Eq.(\ref{eq:K-dot}) can be modified by means of the first 
constraint equation (Hamiltonian constraint) in Eq.(\ref{eq:constraints}). 
In particular, the following two forms would be useful for our purposes.
First, by eliminating  $\mbox{\boldmath $R$}$ from Eq.(\ref{eq:K-dot}), we get 
\begin{equation}
\dot{K} =-\frac{1}{2\a}\frac{D-3}{D-2}(\rho + \frac{D-1}{D-3}p) 
          - K \cdot K +\frac{2}{D-2} \Lambda \ \ . 
\label{eq:K-dot2}                        
\end{equation}
Second, by eliminating the term $K \cdot K$ from Eq.(\ref{eq:K-dot}), we get
\begin{equation}
\dot{K} =\frac{1}{2\a}\frac{D-1}{D-2}(\rho - p) 
          - K^2 - \mbox{\boldmath $R$}  +\frac{2(D-1)}{D-2} \Lambda \ \ . 
\label{eq:K-dot3}                        
\end{equation}

Thus, we obtain the equation for $(\dot{K})_l$ based on Eq.(\ref{eq:K-dot2}),  
\begin{equation}
(\dot{K})_l=-\frac{1}{2\a}\frac{D-3}{D-2}(\rho_l + \frac{D-1}{D-3}p_l)
                -\frac{1}{D-1}\sum_{l'l''}K_{l'}K_{l''}(l'\ l''\ l) 
                +\frac{2 \Lambda \sqrt{V}}{D-2}\d_{l0}
                -(\e\cdot\e)_l\ \ .
\label{eq:(K-dot)l1}                        
\end{equation}
In the same manner, based on Eq.(\ref{eq:K-dot3}), we get 
\begin{equation}
(\dot{K})_l = \frac{1}{2\a}\frac{(D-1)}{(D-2)}
             (\rho_l - p_l)-\sum_{l'l''}K_{l'}K_{l''}(l'\ l''\ l) 
            \ \     +\frac{2 (D-1) \Lambda \sqrt{V}}{D-2}\d_{l0}
                -\mbox{\boldmath $R$}_l \ \ . 
\label{eq:(K-dot)l2}                        
\end{equation}

We also note that, from Eqs.(\ref{eq:<gdg>0}), 
(\ref{eq:average}) and (\ref{eq:A-2}),  
\begin{equation}
\frac{\dot{V}}{V}=K_0 / \sqrt{V}=K_{\rm av} \ \ . 
\label{eq:Vdot}
\end{equation}

The first constraint equation in Eq.(\ref{eq:constraints}) is translated into  
\begin{equation}
\mbox{\boldmath $R$}_l + \frac{D-2}{D-1} \sum_{l'l''}K_{l'}K_{l''}(l'\ l''\ l)
      -\frac{1}{\a} \rho_l -2\Lambda \sqrt{V}\d_{0l} -(\e \cdot \e)_l =0\ \ .
\label{eq:Hamiltonian-constraint}        
\end{equation}

We also note that, taking the inner-product 
with $\frac{1}{\l_l} D_a f_l$ ($l \neq 0$), the Bianchi identity
$D^b \mbox{\boldmath $R$}_{ab} -\frac{1}{2}D_a \mbox{\boldmath $R$} =0 $
turns to  
\begin{equation}
\mbox{\boldmath $R$}_l 
       + \frac{2(D-1)}{(D-3)}\frac{1}{\l_l}(D^aD^b r_{ab})_l=0
                         \ \ (l\neq 0)\ \  .  
\label{eq:Bianchi}        
\end{equation}

Now, taking the inner-product with $\frac{1}{\l_l} D_a f_l$ ($l \neq 0$), 
 the second constraint equation in Eq.(\ref{eq:constraints}) 
 (momentum constraint) turns to 
\begin{equation}
K_l 
  + \frac{D-1}{D-2}\frac{1}{\l_l} \left\{
              (D^a D^b \e_{ab})_l + \frac{1}{2\a}(\vec{D}\cdot \vec{J})_l 
                                   \right\} =0 \ \ (l\neq 0)\ \  .
\label{eq:momentum-constraint}        
\end{equation}

Finally, noting that 
\[
\dot{\mbox{\boldmath $R$}}
= -\mbox{\boldmath $R$}^{ab}\dot{h}_{ab} 
   + D^a\left(D^b \dot{h}_{ab} - D_a(h^{cd}\dot{h}_{cd})\right)\ \ ,
\]
we get  
\begin{equation}
\dot{\mbox{\boldmath $R$}}_l 
=\frac{D-3}{D-1}\sum_{l'l''}K_{l'}\mbox{\boldmath $R$}_{l''}(l'\ l''\ l)
 + \sum_{l'}\mbox{\boldmath $R$}_{l'}\mu_{l'l} \nonumber 
  -\frac{1}{\a}(\vec{D}\cdot\vec{J})_l - 2(\e \cdot r)_l \ \ ,
\label{eq:Rl-dot}                        
\end{equation}
where we used the momentum constraint Eq.(\ref{eq:momentum-constraint}) 
to reach the final form. 

\subsection{Evolution equation for $\<\e_{ab} {\>_{}}_{l,mn}$ 
                 and $\< r_{ab} {\>_{}}_{l,mn}$}

First, we  derive the  evolution equations for $\e_{ab}$ and $r_{ab}$ 
($N=1$, $N_a=0$), 
\begin{eqnarray}
\dot{\e}_{ab} &=& 
     \frac{1}{2\a} (S_{ab}-p h_{ab}) 
     - \frac{D-3}{D-1}K\e_{ab} -r_{ab} +2 (\e \cdot \e)_{ab} \ \ , 
\label{eq:e-dot} \\
\dot{r}_{ab} &=&
     -\frac{1}{D-1}\left\{\Delta K h_{ab}+(D-3) D_aD_b K \right\}  
     +\frac{1}{\a}\frac{1}{D-1} \vec{D}\cdot \vec{J} h_{ab} 
        \nonumber                \\
   &&  -\Delta \e_{ab} -\frac{2}{D-1}\mbox{\boldmath $R$} \e_{ab} 
     +2  D^c D_{(a} \e_{b)c}
     +\frac{2}{D-1}  \e \cdot r h_{ab} \ \ . 
\label{eq:r-dot}     
\end{eqnarray}

From Eq.(\ref{eq:e-dot}) along with Eqs.(\ref{eq:df}) and (\ref{eq:d<Aab>}),    
we get  
\begin{eqnarray}
\<\e_{ab} \dot{{\>_{}}}_{l,mn}&=&
\frac{1}{2\a}\< (S_{ab}-ph_{ab})  {\>_{}}_{l,mn}
 -\left(\ln\sqrt{\l_m\l_n}\right)^\cdot \<\e_{ab} {\>_{}}_{l,mn} 
 -\frac{2}{D-1}\sum_{l'l''}K_{l'}(l'\ l\ l'') \<\e_{ab} {\>_{}}_{l'',mn}
                     \nonumber         \\
 && + \sum_k \sqrt{\frac{\l_k}{\l_n}}\<\e_{ab} {\>_{}}_{l,mk} \mu_{kn} 
   + \sum_k \sqrt{\frac{\l_k}{\l_m}}\<\e_{ab} {\>_{}}_{l,nk} \mu_{km} 
                     \nonumber         \\
 && + \sum_k  \<\e_{ab} {\>_{}}_{k,mn} \mu_{kl} 
 - \< r_{ab} {\>_{}}_{l,mn} 
               -2\< (\e \cdot \e)_{ab}  {\>_{}}_{l,mn} \ \ .
\label{eq:<e>-dot}               
\end{eqnarray}

In the same manner,  we get from Eq.(\ref{eq:r-dot}), 
\begin{eqnarray}
\<r_{ab} \dot{{\>_{}}}_{l,mn}&=& 
 \< \dot{r}_{ab}  {\>_{}}_{l,mn}
 -\left(\ln\sqrt{\l_m \l_n} \right)^\cdot \<r_{ab} {\>_{}}_{l,mn} 
 -\frac{5-D}{D-1}\sum_{l'l''}K_{l'}(l'\ l\ l'') \< r_{ab} {\>_{}}_{l'',mn}
                     \nonumber         \\
 && + \sum_k \sqrt{\frac{\l_k}{\l_n}}\<r_{ab} {\>_{}}_{l,mk} \mu_{kn} 
 + \sum_k \sqrt{\frac{\l_k}{\l_m}}\<r_{ab} {\>_{}}_{l,nk} \mu_{km} 
                     \nonumber         \\
 && + \sum_k  \< r_{ab} {\>_{}}_{k,mn} \mu_{kl} 
               -4\< \e^c_{\ (a} r_{b)c}  {\>_{}}_{l,mn} \ \ ,
\label{eq:<r>-dot}     
\end{eqnarray}
where
\begin{eqnarray}
 \< \dot{r}_{ab}  {\>_{}}_{l,mn}
 &=& \frac{1}{D-1}\sum_{l'l''}\l_{l'}K_{l'}(l'\ l\ l'') 
 \< h_{ab} {\>_{}}_{l'',mn}
         -\frac{D-3}{D-1}\sum_{l'} K_{l'} \< D_aD_b f_{l'}  {\>_{}}_{l,mn} 
                     \nonumber         \\
    &&     +\frac{1}{\a}\frac{1}{D-1}\sum_{l'l''}
     (\vec{D}\cdot \vec{J})_{l'}(l'\ l\ l'')\<  h_{ab} {\>_{}}_{l'',mn}
     -\frac{2}{D-1}\sum_{l'l''}\mbox{\boldmath $R$}_{l'}(l'\ l\ l'')
        \< \e_{ab}{\>_{}}_{l'',mn}
                     \nonumber         \\
   && - \< \Delta \e_{ab} {\>_{}}_{l,mn}
    +2 \< D^c D_{(a} \e_{b)c} {\>_{}}_{l,mn}
                     \nonumber         \\
    && +\frac{2}{D-1}\sum_{l'l''}(\e \cdot r)_{l'}(l'\ l\ l'')
       \<  h_{ab}  {\>_{}}_{l'',mn}\ \ .
\label{eq:<r-dot>}       
\end{eqnarray}

For convenience, we also present the formulas Eq.(\ref{eq:<e>-dot}) and 
Eq.(\ref{eq:<r>-dot}) especially for $l=0$: 
\begin{eqnarray}
 \<\e_{ab} \dot{{\>_{}}}_{mn}&=&
\frac{1}{2\a}\< (S_{ab}-ph_{ab})  {\>_{}}_{mn}
 -\left(\ln\sqrt{\l_m\l_n}\right)^\cdot \<\e_{ab} {\>_{}}_{mn} 
 -\frac{2}{D-1}\sum_{k}K_{k} \<\e_{ab} {\>_{}}_{k,mn}
                     \nonumber         \\
  && + \sum_k \sqrt{\frac{\l_k}{\l_n}}\<\e_{ab} {\>_{}}_{mk} \mu_{kn} 
   + \sum_k \sqrt{\frac{\l_k}{\l_m}}\<\e_{ab} {\>_{}}_{nk} \mu_{km} 
 - \< r_{ab} {\>_{}}_{mn}
                     \nonumber         \\ 
              && -2\< (\e \cdot \e)_{ab}  {\>_{}}_{mn} \ .
\label{eq:<e>-dot2}               
\end{eqnarray}

\begin{eqnarray}
 \<r_{ab} \dot{{\>_{}}}_{mn}&=&
 \< \dot{r}_{ab}  {\>_{}}_{mn}
 -\left(\ln\sqrt{\l_m \l_n} \right)^\cdot \<r_{ab} {\>_{}}_{mn} 
 -\frac{5-D}{D-1}\sum_{k}K_{k} \< r_{ab} {\>_{}}_{k,mn}
                     \nonumber         \\
 && + \sum_k \sqrt{\frac{\l_k}{\l_n}}\<r_{ab} {\>_{}}_{mk} \mu_{kn} 
 + \sum_k \sqrt{\frac{\l_k}{\l_m}}\<r_{ab} {\>_{}}_{nk} \mu_{km}
                     \nonumber         \\ 
              && -4\< \e^c_{\ (a} r_{b)c}  {\>_{}}_{mn} \ \ ,
\label{eq:<r>-dot2}     
\end{eqnarray}
where
\begin{eqnarray}
  \< \dot{r}_{ab}  {\>_{}}_{mn}
 &=& \frac{1}{D-1}\sum_{k} \l_k K_{k} \< h_{ab} {\>_{}}_{k,mn}
              -\frac{D-3}{D-1}\sum_{k} K_{k} \< D_aD_b f_{k}  {\>_{}}_{mn} 
                     \nonumber         \\
    &&     +\frac{1}{\a}\frac{1}{D-1}\sum_{k}
     (\vec{D}\cdot \vec{J})_{k} \<  h_{ab} {\>_{}}_{k,mn}
     -\frac{2}{D-1}\sum_{k}\mbox{\boldmath $R$}_{k}
        \< \e_{ab}{\>_{}}_{k,mn}
                     \nonumber         \\
   && - \< \Delta \e_{ab} {\>_{}}_{mn}
    +2 \< D^c D_{(a} \e_{b)c} {\>_{}}_{mn}
     +\frac{2}{D-1}\sum_{k}(\e \cdot r)_{k}
       \<  h_{ab}  {\>_{}}_{k,mn}\ \ .
\label{eq:<r-dot>2}
\end{eqnarray}
  Eqs. (\ref{eq:dlambda-2})-(\ref{eq:Kl-dot}), (\ref{eq:(K-dot)l1}) 
  (or (\ref{eq:(K-dot)l2})), (\ref{eq:Vdot})-(\ref{eq:Rl-dot}),  
  and  (\ref{eq:<e>-dot})-(\ref{eq:<r-dot>}) are 
  the fundamental equations for the  investigation of 
   the spectral evolution of the Universe. 

They form hierarchy equations and we can continue to get equations for 
higher hierarchies. This hierarchical property 
 is a reasonable consequence since we are   
looking at the global quantities, and not the local ones.  
 In practical applications, thus, 
we need to make a suitable truncation.    
Typically, we get equations of higher order in $\e_{ab}$  and  $r_{ab}$ 
when  we continue to go into further hierarchies. 
We can often regard $\e_{ab}$, 
$r_{ab}$,  and their spatial derivatives are small in cosmological 
applications. In such cases, the truncation becomes a reasonable 
approximation procedure.

\section{Basic applications of the spectral equations}
\label{section:IV}

\subsection{The Friedman-Robertson-Walker Universe}
\label{subsection:IV-1}

As a basic example, let us consider a closed 
Friedmann-Robertson-Walker  Universe. 

We set $\e_{ab}=0$ and $r_{ab}=0$. We also set 
$\vec{J}=0$. From Eqs. (\ref{eq:Bianchi}) and
 (\ref{eq:momentum-constraint}), 
we get $\mbox{\boldmath $R$}_l=0$ and $K_l=0$ ($l\neq 0$). 
Then, we  get $\rho_l=0$ ($l\neq 0$) from 
Eq.(\ref{eq:Hamiltonian-constraint}), noting that 
$(0\ 0\ l)=\frac{1}{\sqrt{V}}\d_{0l}$.  
Eq.(\ref{eq:Kl-dot}) along with Eq.(\ref{eq:(K-dot)l1}) 
imply  that $p_l=0$ ($l\neq 0$), noting that
 $\mu_{mn}= - \frac{1}{2}\d_{mn} K_0/\sqrt{V}$  
in the present case (Eq.(\ref{eq:mu_mn})). 

Now, Eq.(\ref{eq:dlambda-2}) with Eq.(\ref{eq:Vdot}) yield
\[
\dot{\l}_n= - \frac{2}{D-1}\l_n \frac{\dot{V}}{V}\ \ , 
\]
thus, 
\[
\l_n(t)={\left(\frac{V(0)}{V(t)}\right)}^{\frac{2}{D-1}}\l_n (0)\ \ .
\]
It is a simple scaling behavior expected from  the dimensionality of $\l_n$. 

Noting that $\mu_{l0}= - \frac{1}{2}\d_{l0} K_0/\sqrt{V}$ 
(Eq.(\ref{eq:mu_mn})),  
we get from Eq.(\ref{eq:Rl-dot}) for $l=0$ with 
Eq.(\ref{eq:Vdot}), 
\[
\dot{\mbox{\boldmath $R$}}_0
= -\frac{5-D}{2(D-1)}  
\frac{\dot{V}}{V} \mbox{\boldmath $R$}_0 \ \ .
\]
Thus, 
\[
\mbox{\boldmath $R$}_0(t)
= \left(\frac{V(0)}{V(t)}\right)^{\frac{5-D}{2(D-1)}}
\mbox{\boldmath $R$}_0(0)\ \ ,
\]
or, noting Eq.(\ref{eq:average}),  
\[
\mbox{\boldmath $R$}_{\rm av}(t)
= \left(\frac{V(0)}{V(t)}\right)^{\frac{2}{D-1}}
\mbox{\boldmath $R$}_{\rm av}(0)\ \ .
\]
It is also a simple scaling behavior expected from the 
dimensionality of $\mbox{\boldmath $R$}$.  

On the other hand, 
from Eq.(\ref{eq:Hamiltonian-constraint}) for $l=0$ with 
Eq.(\ref{eq:Vdot}), we get
\[
\frac{D-2}{D-1} \left(\frac{\dot{V}}{V}\right)^2 
+ \mbox{\boldmath $R$}_{\rm av} 
-\frac{1}{\a} \rho_{\rm av} - 2\Lambda = 0\ \ , 
\]
or in a more familiar form,
\[
\left(\frac{\dot{a}}{a}\right)^2 + \frac{k}{a^2} 
- \frac{1}{(D-1)(D-2)}\frac{1}{\a} \rho_{\rm av} 
- \frac{2}{(D-1)(D-2)}\Lambda =0 \ \ ,
\]
where we introduced the scale factor $a$ as $V \propto a^{D-1}$ and 
the curvature index $k:=\mbox{\boldmath $R$}_{\rm av}(0)a(0)^2$.  

It can be solved once the  matter content is specified.

\subsection{A universal formula for the spectral distance between 
two  Universes that are geometrically close to each other}  
\label{subsection:IV-2}

The spectral evolution equations developed in the previous sections 
are of essential significance 
for the study of spacetime physics along the line of 
the spectral representation of geometrical structures.

We have the spectral distance $d_N ({\cal G},{\cal G}')$ 
as a measure of  closeness between two spatial geometries $\cal G$ and 
  ${\cal G}'$~\cite{MS-spectral}. 
  It measures the difference between $\cal G$ and 
  ${\cal G}'$ as the difference of ``sounds" of them, i.e. 
  the difference of the spectra. 

Then we are given a basic arena for spacetime physics,  i.e. 
the  ``space of all spaces" ${\cal S}_N$; it is basically 
a space of all compact Riemannian geometries equipped with 
$d_N ({\cal G},{\cal G}')$~\cite{MS-space}.   

Since we now have the spectral evolution equations developed in the 
previous sections, we can  investigate the time evolution of 
$d_N ({\cal G},{\cal G}')$. The fundamental importance of such an 
investigation becomes clear when we take $\cal G$ as the real Universe and 
${\cal G}'$ as a model Universe which is expected to be ``close" to 
$\cal G$~\cite{MS-JGRG,MS-AVE}. 
There is no guarantee that the model Universe remains a good model for the 
real Universe in the future also~\cite{AVE,AVE2}. 
Stated in terms of the spectral distance, 
there is no guarantee that   
$d_N ({\cal G},{\cal G}')$  remains small in the future.    
The detailed investigations along this line 
would be done separately. Here we only look at some basic features of 
the evolution of $d_N ({\cal G},{\cal G}')$ when $\cal G$ and 
  ${\cal G}'$ are very close initially in ${\cal S}_N$. 

In this subsection, we derive a universal formula for 
$d_N ({\cal G},{\cal G}')$ when $\cal G$ and ${\cal G}'$ are 
geometrically close in ${\cal S}_N$, which is valid independently of 
the detailed form of $d_N ({\cal G},{\cal G}')$ nor the gravity theory. 
In the next subsection, we discuss its time evolution.  

First, it is appropriate to  recall basic settings of 
the spectral representation~\cite{MS-spectral}. 

  We set the eigenvalue problem on each manifold 
  $\cal G$ and ${\cal G}'$, 
  \[
  \Delta f =-\l f \ \ ,
  \]  
  then the set of eigenvalues (numbered in increasing order) is obtained;  
  $\{\l_m \}_{m=0}^\infty$ for $\cal G$  and 
  $\{\l'_n \}_{n=0}^\infty$ for ${\cal G}'$.  

  Now the spectral distance between $\cal G$ and 
  ${\cal G}'$ is defined as~\cite{MS-space,MS-AVE}  
\begin{equation}
  d_N ({\cal G},{\cal G}')= \sum_{n=1}^N {\cal F} 
  \left( \frac{\l'_n}{\l_n} \right)\ \ ,
\label{eq:d_N_general}
\end{equation}
  where ${\cal F}(x)$ ($x>0$) is a smooth function 
  which satisfies   ${\cal F} \geq 0$,    
  ${\cal F}(1/x)={\cal F}(x)$, 
  ${\cal F}(y)>{\cal F}(x)$ if  $y > x \geq 1$ and 
  ${\cal F}(1)=0$. Then,  it follows that ${\cal F}''(1) \geq 0$.  However, 
   in order  to let $d_N ({\cal G},{\cal G}')$ 
  detect a fine difference between ${\cal G}$ and ${\cal G}'$ when 
  they are very close to each other  in ${\cal S}_N$,  
we further postulate that ${\cal F}''(1)>0$.  (See Eq. (\ref{eq:d_N-close}).)

In particular, it is convenient to choose as ${\cal F}$, 
 ${\cal F}_1(x)=\frac{1}{2} \ln \frac{1}{2}(\sqrt{x}+1/\sqrt{x}) $.  
Then  we get~\cite{MS-spectral}
\begin{equation}
d_N ({\cal G},{\cal G}')
=\frac{1}{2} \sum_{n=1}^N \ln \frac{1}{2}
\left(
\sqrt{\frac{\l_n'}{\l_n}}
+\sqrt{\frac{\l_n}{\l'_n}}
\right)\ \ .
\label{eq:d_N} 
\end{equation}
It turns out that $d_N ({\cal G},{\cal G}')$ does not satisfy the 
triangle inequality. However, it does not cause a serious problem.
Let  $Riem$ be  the space
 of all $(D-1)$-dimensional, compact Riemannian geometries without 
 boundaries. Then it is proved that the space $(Riem, d_N)/_\sim $,  
 where $d_N$ is given by Eq.(\ref{eq:d_N}), is a metrizable 
 space~\cite{MS-space}. 
 (Here $_\sim$ indicates identification of isospectral manifolds~\cite{CH}.)  
 It justifies to regard 
$d_N ({\cal G},{\cal G}')$ as a distance, provided that we are careful
when the triangle inequality matters in the argument.   
 
 The above property  is shown  in the  following way. 
Since the breakdown of the triangle inequality turns out to be a mild one 
in a certain sense~\cite{MS-space}, 
it is expected that one can  make a slight modification of ${\cal F}_1$ 
to recover the inequality. 
Indeed  we can find ${\cal F}_0(x):=\frac{1}{2}\ln\max(\sqrt{x},1/\sqrt{x})$ as 
a  modification of ${\cal F}_1$. 
In this case,  Eq.(\ref{eq:d_N_general}) becomes 
\[
\bar{d}_N({\cal G}, {\cal G}')
=\frac{1}{2}\sum_{n=1}^{N}
\ln\max\left(\sqrt{\frac{\l_n'}{\l_n}},
           \sqrt{\frac{\l_n}{\l_n'}} \right) \ \ . 
\]
It is easy to show that $\bar{d}_N({\cal G}, {\cal G}')$ satisfies 
all of the axioms of distance, so that it is a distance. 
Now, one can show that $(Riem, d_N)/_\sim $ and $(Riem, \bar{d}_N)/_\sim $ 
are homeomorphic to each other. Thus,  $(Riem, d_N)/_\sim $ is a metrizable 
space since  $(Riem, \bar{d}_N)/_\sim $ is a metric space. 
Let ${\cal S}_N$  be a completion of $(Riem, d_N)/_\sim $. 
The definition for $d_N$ has a  more convenient form than 
the one for $\bar{d}_N$ since 
the latter includes $max$ symbol. Thus, in the actual applications, 
$d_N$ is more useful than $\bar{d}_N$. On the other hand, 
when we need to discuss  mathematical properties of ${\cal S}_N$ precisely,   
$\bar{d}_N$ is appropriate.

 Let us consider the situation that  ${\cal G}$ and ${\cal G}'$ possess the 
 same topological structure and that they are very close in ${\cal S}_N$. 
One can imagine that  ${\cal G}'=(\Sigma, h')$  represents the real Universe 
at some instant of time, and ${\cal G}=(\Sigma, h)$ is a model
 Universe corresponding to ${\cal G}'=(\Sigma, h')$. 
 We introduce the difference in the spatial metric
\begin{equation}
\gamma_{ab}:=h'_{ab}-h_{ab} \ \ ,  
\label{eq:gamma}
\end{equation} 
and  we treat $\gamma_{ab}$ as a  small quantity. 
 
We regard the model Universe as a reference 
point, based on which we  evaluate several quantities. In particular, 
we make use of the time-slicings of the model and investigate 
the time evolutions with respect to them.  

Now, from   Eq.(\ref{eq:dloglambda}), we get   
\begin{equation}
\d \ln \l_n =\frac{\l_n'-\l_n}{\l_n}
= -\<\overline{\gamma}_{ab}{\>_{}}_{n} 
                -\frac{1}{2}\<\gamma {\>_{}}_{n} \ \ , 
\label{eq:dloglambda-distance}
\end{equation} 
where $\gamma:=h^{ab}\gamma_{ab}$.

Looking at Eq.(\ref{eq:d_N_general}), we see that 
\[
{\cal F} \left( \frac{\l_n'}{\l_n} \right)
= {\cal F} \left( 1-\left(\<\overline{\gamma}_{ab}{\>_{}}_{n} 
                +\frac{1}{2}\<\gamma {\>_{}}_{n}\right) \right)
= \frac{1}{2}{\cal F}''(1) 
    \left(\<\overline{\gamma}_{ab}{\>_{}}_{n}
          +\frac{1}{2}\<\gamma {\>_{}}_{n}\right)^2 + O(\varepsilon^3) \ \ . 
\]
Here  we note that ${\cal F}(1)={\cal F}'(1)=0$, and $\varepsilon$ 
indicates a small quantity in the same order as $\gamma$.     

Leaving only the leading term,  we thus get 
\begin{equation}
d_N ({\cal G},{\cal G}')
= \frac{1}{2}{\cal F}''(1)  \sum_{n=1}^{N} 
    \left(\<\overline{\gamma}_{ab}{\>_{}}_{n} + 
              \frac{1}{2}\<\gamma {\>_{}}_{n}\right)^2 \ \ . 
\label{eq:d_N-close}
\end{equation}
Here we note the  postulation  ${\cal F}''(1) > 0$. 

It would be also helpful to view Eq.(\ref{eq:d_N-close}) as
\begin{equation}
d_N ({\cal G},{\cal G}')
 =\frac{1}{2}{\cal F}''(1) 
        \vec{\mbox{\boldmath $\gamma$}}
                     \cdot\vec{\mbox{\boldmath $\gamma$}} \ \ ,
\label{eq:d_N-close2}
\end{equation}
where $\vec{\mbox{\boldmath $\gamma$}}$ is a vector in 
$\mbox{\boldmath $R$}^N$ whose $n$-th component is 
$\<\overline{\gamma}_{ab}{\>_{}}_{n} + \frac{1}{2}\<\gamma {\>_{}}_{n}$,  and 
a standard Euclidean inner-product is implied.

 Hence,  we get a universal result on 
$d_N ({\cal G},{\cal G}')$ when $\cal G$ and ${\cal G}'$ are very close 
in ${\cal S}_N$: {\it The leading behavior of the 
spectral distance $d_N ({\cal G},{\cal G}')$ is given by 
Eq.(\ref{eq:d_N-close}) (or Eq.(\ref{eq:d_N-close2})),  
 irrespective of the detailed form of the spectral distance nor 
 of the gravity theory.}

In the case of Eq.(\ref{eq:d_N}), we have chosen  as $\cal F$,  
${\cal F}_1(x)=\frac{1}{2} \ln \frac{1}{2}(\sqrt{x}+1/\sqrt{x}) $, 
hence ${\cal F}''(1)=\frac{1}{8}$. Thus, we get 
\begin{equation}
d_N ({\cal G},{\cal G}')
= \frac{1}{16}  \sum_{n=1}^{N} 
    \left(\<\overline{\gamma}_{ab}{\>_{}}_{n} + 
              \frac{1}{2}\<\gamma {\>_{}}_{n}\right)^2 \ \ ,  
\label{eq:d_N-close3}
\end{equation}
or 
\begin{equation}
d_N ({\cal G},{\cal G}')
 =\frac{1}{16} 
        \vec{\mbox{\boldmath $\gamma$}}
                     \cdot\vec{\mbox{\boldmath $\gamma$}} \ \ ,
\label{eq:d_N-close4}
\end{equation}
where  $\vec{\mbox{\boldmath $\gamma$}}$ is the same vector as in 
Eq.(\ref{eq:d_N-close2}).

\subsection{Time evolution of a small geometrical discrepancy 
between the real and a model Universes} 
\label{subsection:IV-3}

Here we investigate the time evolution of the spectral distance 
 for the two `nearby' Universes as described in the 
previous subsection.  Taking the time derivative of the both sides of 
Eq.(\ref{eq:d_N}), we get 
\begin{equation}
\dot{d}_N ({\cal G},{\cal G}') = 
\frac{1}{4}\sum_{n=1}^N \frac{\frac{\l'_n}{\l_n}-1}{\frac{\l'_n}{\l_n}+1} 
\left(\ln \frac{\l'_n}{\l_n}\right)^\cdot \ \ .
\label{eq:d_N-dot1}
\end{equation}

Now Eq.(\ref{eq:dlambda-2}) can be represented as
\[
(\ln \l_n)^\cdot= -\frac{2}{D-1}
            \sum_{l} (1 + \frac{D-3}{4}\frac{\l_l}{\l_n})\ K_l \  (l\ n\ n) 
              - 2 \<\e_{ab} {\>_{}}_{n}  \ \ .
\] 
In the cosmological problems, it is often useful to separate the term for 
$l=0$ from the terms for $l\geq 1$ in the summation like in the above 
equation: This separation is useful when the Universe is described 
by a  homogeneous geometry plus small perturbations. 
Noting  Eq.(\ref{eq:Vdot}), 
we thus get
\begin{equation}
(\ln \l_n)^\cdot = -2\left(\frac{1}{D-1}\frac{\dot{V}}{V} 
            +\frac{1}{D-1} \<K- K_{\rm av}{\>_{}}_{n} 
            -\frac{D-3}{4(D-1)}\frac{1}{\l_n} \< \Delta K {\>_{}}_{n} 
           + \< \e_{ab} {\>_{}}_{n} \right) \ \ . 
\label{eq:lnlambda_n-dot}
\end{equation}
Let us define 
\begin{eqnarray}
H:&=&\frac{1}{D-1} \frac{\dot{V}}{V} \ \ , \nonumber \\
\iota_n:&=&\frac{1}{D-1} \sum_{l \geq 1} 
             (1 + \frac{D-3}{4}\frac{\l_l}{\l_n})\ K_l \  
                         (l\ n\ n) \nonumber  \\
        &=& \frac{1}{D-1} \<K- K_{\rm av}{\>_{}}_{n} 
             -\frac{D-3}{4(D-1)}\frac{1}{\l_n} \< \Delta K {\>_{}}_{n} 
                                              \ \ ,   
\label{eq:parameters}                         \\    
\a_n:&=& \< \e_{ab} {\>_{}}_{n} \ \ . \nonumber
\end{eqnarray}
Here  $H$ is an analogous quantity to the Hubble constant; $\iota_n$ 
is attributed to the inhomogeneity while $\a_n$ is to the anisotropy 
of the geometry. 
 Then, the formula (\ref{eq:lnlambda_n-dot}) can be 
 represented more concisely as 
\begin{equation}
(\ln \l_n)^\cdot=-2H_n \ \ ,
\label{eq:lnlambda_n-dot2}
\end{equation}
where 
\begin{equation}
H_n:= H + \iota_n + \alpha_n \ \ .
\label{eq:Hubble_n}
\end{equation}
The quantity $H_n$  can be interpreted as the effective Hubble constant 
observed at the scale 
$\l_n^{-1/2}$ since it determines the rate of change of $\l_n$. Thus, 
Eq.(\ref{eq:Hubble_n}) 
describes the modification of the {\it effective} Hubble parameter 
{\it at the scale} $\l_n^{-1/2}$  due to 
 inhomogeneity and  anisotropy  of the Universe {\it at that scale}.

Leaving only the leading terms in Eq.(\ref{eq:d_N-dot1}) with 
the help of Eqs.(\ref{eq:lnlambda_n-dot2}) and (\ref{eq:Hubble_n}), 
we thus get
\begin{equation}
\dot{d}_N({\cal G},{\cal G}')=\frac{1}{4}\vec{\mbox{\boldmath $\gamma$}}
\cdot \d \vec{\mbox{\boldmath $H$}}\ \ ,
\label{eq:d_N-dot2}
\end{equation}
where $\d \vec{\mbox{\boldmath $H$}}$ denotes a vector in 
$\mbox{\boldmath $R$}^N$ whose $n$-th component is 
$\d H_n := H'_n - H_n$.\footnote{ 
In this subsection, $\d Q$ denotes the difference in a quantity $Q$ for 
two spaces $\cal G$ and ${\cal G}'$, defined as 
$Q$ for ${\cal G}'$ (the second entry of $d_N(\cdot, \cdot)$) minus 
$Q$ for ${\cal G}$ (the first entry of $d_N(\cdot, \cdot)$). 
Wherever necessary, we employ  the notation $\d \{\  \cdot \  \}$ (rather than 
$\d (\  \cdot \  )$) to avoid any confusion with the  $\d$-function.}
 On the other hand, from Eq.(\ref{eq:lnlambda_n-dot2}) with 
 Eq.(\ref{eq:dloglambda-distance}), we can derive  
 $\dot{\vec{\mbox{\boldmath $\gamma$}}}= 2 \d \vec{\mbox{\boldmath $H$}}$, 
 which is compatible with Eqs.(\ref{eq:d_N-close4}) and (\ref{eq:d_N-dot2}).   

From Eq.(\ref{eq:d_N-dot2}), we get 
\begin{equation}
\ddot{d}_N({\cal G},{\cal G}')=
\frac{1}{2}\d \vec{\mbox{\boldmath $H$}}\cdot 
             \d \vec{\mbox{\boldmath $H$}} 
+ \frac{1}{4}\vec{\mbox{\boldmath $\gamma$}}\cdot 
\d \dot{\vec{\mbox{\boldmath $H$}}}
       \ \ .
\label{eq:d_N-2dot}
\end{equation}

Let us introduce  $q_n:=-(1 + \frac{\dot{H}_n}{H_n^2})$, which can be 
interpreted as  the effective 
 deceleration parameter at the scale $\l_n^{-1/2}$. 
Then, $\dot{H}_n=-(1+q_n)H_n^2$, so that 
\begin{equation}
\d \dot{H}_n= -2(1+q_n)H_n\d H_n - H_n^2\d q_n \ \ .
\label{eq:delta H_n-dot}
\end{equation}

Since $\d \dot{H}_n$ appears in Eq.(\ref{eq:d_N-2dot}) only in the form of 
$\vec{\mbox{\boldmath $\gamma$}}\cdot 
\d \dot{\vec{\mbox{\boldmath $H$}}}$, it suffices to  
 estimate $\d \dot{H}_n$ by leaving only the leading terms  
in Eq.(\ref{eq:delta H_n-dot}). 
First, using Eq.(\ref{eq:Hubble_n}), we note that 
\[
q_n \simeq (1-\frac{2}{H}(\iota_n+\a_n))q 
         -\frac{2}{H}(\iota_n + \a_n) 
          - \frac{1}{H^2}(\dot{\iota_n} + \dot{\a}_n) \ \ ,
\] 
where $q:= -(1 + \frac{\dot{H}}{H^2})$. 

Next, it is also straightforward to get  estimations 
\begin{eqnarray*}
(1+q_n)H_n \d H_n & \simeq &  
 \left[(1+q)\{H - (\iota_n + \a_n)\} -\frac{1}{H}(\dot{\iota_n} + \dot{\a_n}) 
 \right] \d H 
 + (1+q)H \d \{\iota_n + \a_n \}
 \ \ , \\
H_n^2 \d q_n & \simeq & 
H^2 \d q + 2\{(1+q)(\iota_n + \a_n) 
+ \frac{1}{H}(\dot{\iota_n} + \dot{\a_n}) \} \d H  \\
&&  -2(1+q)H \d \{ \iota_n + \a_n \}
- \d \{ \dot{\iota_n} + \dot{\a_n} \} \ \ . 
\end{eqnarray*} 

Looking at  Eq.(\ref{eq:delta H_n-dot}), we thus estimate 
\[
\d \dot{H}_n \simeq -\d \{ (1+q)H^2 \} + \d \{\iota_n + \a_n \dot{\}\ } \ \ .
\]

Finally, we reach the estimation
\begin{equation}
\ddot{d}_N ({\cal G},{\cal G}') \simeq 
\frac{1}{2}\d \vec{\mbox{\boldmath $H$}}\cdot \d \vec{\mbox{\boldmath $H$}}
-\frac{1}{4}(\sum_{n=1}^N \gamma_n) \d \{ (1+q)H^2 \}  
+ \frac{1}{4} \vec{\mbox{\boldmath $\gamma$}}\cdot 
\d \{ \vec{\mbox{\boldmath $\iota$}} + \vec{\mbox{\boldmath $\a$}}\dot{\} \ } 
\ \ , 
\label{eq:ddot d_N}
\end{equation}
where $\gamma_n:=\<\overline{\gamma}_{ab}{\>_{}}_{n} 
+ \frac{1}{2}\<\gamma {\>_{}}_{n}$ is  the $n$-th component of 
$\vec{\mbox{\boldmath $\gamma$}}$, and 
$\vec{\mbox{\boldmath $\iota$}}$ and $\vec{\mbox{\boldmath $\a$}}$ 
are vectors whose $n$-th components 
are $\iota_n$ and $\a_n$, respectively. 
We note once again the definition of   $\d Q$, the difference of $Q$ 
in $\cal G$ and ${\cal G}'$, in this subsection (see the footnote 
just after Eq.(\ref{eq:d_N-dot2})).
The difference in metric, $\gamma_{ab}$, is also defined in the 
same manner (Eq.(\ref{eq:gamma})). Thus, Eq.(\ref{eq:ddot d_N}) is symmetric in 
$\cal G$ and ${\cal G}'$, as it should be. 

 To make a detailed study, we need to investigate  
 $\<\e_{ab} \dot{{\>_{}}}_{n}$ and $\<r_{ab} \dot{{\>_{}}}_{n}$ also. 
   It is helpful  to note that  $\mu_{mn}$ in 
  Eq.(\ref{eq:mu_mn}) gets simplified in the present case as  
\begin{eqnarray}
\mu_{mn}&=& -\frac{1}{2}\frac{\dot{V}}{V} \d_{mn} + O(\varepsilon) \nonumber \\
        &=& - \frac{D-1}{2} H \d_{mn} + O(\varepsilon) \ \ .
\label{eq:mu_mn-distance}                                     
\end{eqnarray}
 
  Now,  we set $m=n$ in Eq.(\ref{eq:<e>-dot2}). 
 First, let us omit the first and the last 
 terms on the R.H.S. of Eq.(\ref{eq:<e>-dot2}).  
 Next, we leave only the $k=0$ part in the third term, which gives 
 $-\frac{2}{D-1} \frac{\dot{V}}{V} \<\e_{ab}{\>_{}}_{n} $
 $=-2H \<\e_{ab} {\>_{}}_{n}$ with the help of 
Eq.(\ref{eq:Vdot}).  
Next, with the help of Eq.(\ref{eq:mu_mn-distance}), 
the fourth and the fifth terms on the R.H.S. in Eq.(\ref{eq:<e>-dot2})
are approximated together as 
$-\frac{\dot{V}}{V} \<\e_{ab}{\>_{}}_{n}$
$=-(D-1)H\<\e_{ab} {\>_{}}_{n}$. Finally, with the help of 
Eqs.(\ref{eq:lnlambda_n-dot2}) and (\ref{eq:Hubble_n}), 
the second term in Eq.(\ref{eq:<e>-dot2}) yields $2H\<\e_{ab}{\>_{}}_{n}$ 
within the order of magnitude discussed now.   
Thus, we estimate
\begin{equation}
\<\e_{ab} \dot{{\>_{}}}_{n} 
\simeq -(D-1)H \<\e_{ab} {\>_{}}_{n} - \<r_{ab}{\>_{}}_{n}\ \ .      
\label{eq:<e>-dot-approx}
\end{equation}
In a similar manner, we can estimate $\<r_{ab} \dot{{\>_{}}}_{n}$ 
by  Eqs.(\ref{eq:<r>-dot2}) and (\ref{eq:<r-dot>2}). 
We set $m=n$ in these equations. Among all the terms 
in  Eq.(\ref{eq:<r-dot>2}), it is a reasonable approximation  
to leave only the $k=0$ 
part of the fourth term on the R.H.S.,  which gives 
$-\frac{2}{D-1} \frac{\mbox{\boldmath $R$}_0}{\sqrt{V}}
\<\e_{ab}{\>_{}}_{n}$. (This approximation is especially effective when 
we discuss  the large-scale behaviors (low-lying spectra)  such as 
$\l_n^{-1/2}>> c_s \tau$, where $c_s$ is the sound velocity of the matter and 
$\tau$ is the typical cosmological time-scale of interest.) 
In Eq.(\ref{eq:<r>-dot2}), we omit the last term 
on the R.H.S. The third term can be estimated as 
$-\frac{5-D}{D-1} \frac{\dot{V}}{V} \<r_{ab}{\>_{}}_{n}$
$=-(5-D)H\<r_{ab}{\>_{}}_{n}$, while the 
fourth and the fifth terms together are estimated as 
$- \frac{\dot{V}}{V} \<r_{ab}{\>_{}}_{n}$
$= -(D-1)H\<r_{ab}{\>_{}}_{n}$. Finally, the second term yields 
$2H\<r_{ab}{\>_{}}_{n}$. Thus, we can estimate as
\begin{equation}
\<r_{ab} \dot{{\>_{}}}_{n} \simeq
 -2H \<r_{ab}{\>_{}}_{n} 
-\frac{2}{D-1} \mbox{\boldmath $R$}_{\rm av}\<\e_{ab}{\>_{}}_{n}\ \ .
\label{eq:<r>-dot-approx}
\end{equation}

Now, we can estimate $\ddot{d}_N ({\cal G},{\cal G}')$ further based on  
 Eq.(\ref{eq:ddot d_N}). First, $\dot{\alpha_n}$ is given by 
 Eq.(\ref{eq:<e>-dot-approx}), 
\begin{equation}
\dot{\alpha_n} \simeq -(D-1)H \alpha_n - \<r_{ab}{\>_{}}_{n}\ \ .
\label{eq:alpha-dot}
\end{equation} 
Second, regarding $\dot{\iota_n}$, we go back to the definition 
of $\iota_n$ (Eq.(\ref{eq:parameters})).   
With the help of Eqs.(\ref{eq:d(lmn)}), (\ref{eq:Kl-dot}), 
(\ref{eq:(K-dot)l1}), (\ref{eq:lnlambda_n-dot2})
 and (\ref{eq:mu_mn-distance}), it is straightforward to  get estimations
\begin{eqnarray*}
\left(\frac{\l_l}{\l_n}\right)^\cdot &=& O(\varepsilon)\ \ , \\
(l\  n\  n \dot{)} &=& -\frac{D-1}{2}H (l\ n\ n) + O(\varepsilon)\ \ , \\
\dot{K}_l &=& -\frac{5-D}{2}HK_l 
      -\frac{D-3}{2(D-2)}\frac{1}{\a}(\rho_l + \frac{D-1}{D-3}p_l) 
       + O(\varepsilon^2) \ \ .       
\end{eqnarray*}
Thus we get 
\begin{equation}
\dot{\iota_n} \simeq -2H \iota_n -\frac{D-3}{2(D-2)}\frac{1}{\a} {\cal M}_n\ \ ,
\label{eq:iota-dot}
\end{equation} 
where 
\[
{\cal M}_n := 
 \frac{1}{D-1}\{ \<\rho - \rho_{\rm av}{\>_{}}_{n} + 
                   \frac{D-1}{D-3}\<p - p_{\rm av}{\>_{}}_{n} \}
             -\frac{D-3}{4(D-1)}\frac{1}{\l_n} 
             \{ \< \Delta \rho {\>_{}}_{n} 
             +\frac{D-1}{D-3} \< \Delta p {\>_{}}_{n} \} \ \ .
\]
With the help of Eqs.(\ref{eq:alpha-dot}) and (\ref{eq:iota-dot}), 
Eq.(\ref{eq:ddot d_N}) can be expressed  as 
\begin{eqnarray}
\ddot{d}_N ({\cal G},{\cal G}') \simeq && 
\frac{1}{2}\d \vec{\mbox{\boldmath $H$}}\cdot \d \vec{\mbox{\boldmath $H$}}
-\frac{1}{4}(\sum_{n=1}^N \gamma_n) \d \{(1+q)H^2 \} \nonumber \\
&& -\frac{1}{4} \vec{\mbox{\boldmath $\gamma$}}\cdot 
\d \left\{ 2H \vec{\mbox{\boldmath $\iota$}} 
+(D-1)H \vec{\mbox{\boldmath $\a$}}
+ \vec{\mbox{\boldmath $c$}} 
+\frac{(D-3)}{2(D-2)}\frac{1}{\a} \vec{\mbox{\boldmath ${\cal M}$}}\right\}  
 \ \ , 
\label{eq:ddot d_N2}
\end{eqnarray} 
where $\vec{\mbox{\boldmath $c$}}$ and $\vec{\mbox{\boldmath ${\cal M}$}}$
 are vectors whose $n$-th components are, respectively, 
 $\<r_{ab}{\>_{}}_{n}$ and ${\cal M}_n$. 

Eqs. (\ref{eq:d_N-close4}), (\ref{eq:d_N-dot2}) and (\ref{eq:ddot d_N2}) 
give us an idea on   the  factors  that 
influence the validity of a cosmological model.

First, a particular combination  
$\gamma_n:=\<\overline{\gamma}_{ab}{\>_{}}_{n} 
+ \frac{1}{2}\<\gamma {\>_{}}_{n}$ determines 
the spectral distance 
$d_N({\cal G},{\cal G}')$. Second, 
$\vec{\mbox{\boldmath $\gamma$}}\cdot \d \vec{\mbox{\boldmath $H$}}$ 
determines the rate of change of $d_N({\cal G},{\cal G}')$, 
$\dot{d}_N({\cal G},{\cal G}')$. Whether  $d_N({\cal G},{\cal G}')$ 
 decreases or not is governed  by the relative directions 
 of $\vec{\mbox{\boldmath $\gamma$}}$ and $\d \vec{\mbox{\boldmath $H$}}$ 
 in $\mbox{\boldmath $R$}^N$: One of the criteria for a good cosmological model 
 would be that it makes the quantity 
 $\vec{\mbox{\boldmath $\gamma$}}\cdot \d \vec{\mbox{\boldmath $H$}}$ 
 negative, or non-negative and small at least.  
Third, the acceleration, $\ddot{d}_N({\cal G},{\cal G}')$  
is determined by several factors: 
The difference in the effective Hubble parameter, 
$\d \vec{\mbox{\boldmath $H$}}$, has always a repulsive effect. 
Even though  $\vec{\mbox{\boldmath $\gamma$}}=0$ initially, so that
$d_N=\dot{d}_N=0$ initially, the spectral distance increases if 
$\d \vec{\mbox{\boldmath $H$}}\neq 0$. It is required  
that  the other terms  in 
Eq.(\ref{eq:ddot d_N2}) as a whole  should be negative or at least 
non-negative and small in order to get a good model. 

Let $\tau$ be the typical time-scale with which we want to 
discuss the  evolution of the Universe. Then 
we can summarize the criteria for a good cosmological model as follows:
\begin{description}
\item{(C1)} $d_N({\cal G},{\cal G}')$ is small.
\item{(C2)} $\tau \dot{d}_N({\cal G},{\cal G}')$ is negative, or at least,
 non-negative and small. 
\item{(C3)} $\tau^2 \ddot{d}_N({\cal G},{\cal G}')$ is negative, 
or at least, non-negative and small. 
\end{description}

\section{Discussion}
\label{section:V}

In this paper, we have derived the spectral evolution equations 
of the Universe. A set of these equations forms one of the  essential elements  
of the general scheme of spectral representation~\cite{MS-spectral}:  
Now we have a space of all spaces~\cite{MS-space}, ${\cal S}_N$,  equipped with 
the spectral distance~\cite{MS-spectral}, $d_N$, and the evolution equations 
on ${\cal S}_N$. 
  
 The spectral evolution equations  are expected to be  useful for studying 
 the time evolution of the global geometrical structures of the Universe, 
 since the spectra are especially suitable  for describing global 
 properties of a space. 

The significance of the spectral evolution equations becomes prominent 
in situations when  we need to  handle a set of spaces  
rather than just one space, e.g. 
the comparison between the real Universe and its model, the relation 
between the underlying topological structures and its low energy behavior 
(scale-dependent topology~\cite{Vis,MS-scale}), and so on.

As one of the important applications of the spectral evolution equations,  
we can now investigate a fundamental problem in cosmology: 
Whether cosmology is possible, viz. under what conditions and 
to what extent a model reflects 
the real spacetime faithfully (``The averaging/model-fitting problem 
  in cosmology~\cite{AVE,AVE2}). 
  
  As the first step in this direction, we have investigated in 
  \S \ref{section:IV}  
  the spectral distance between two very close Universes and its time 
  evolution. It is  interesting that the spectral distance in this situation  
  is universally determined by the quantity 
  $\gamma_n:=$
  $\<\overline{\gamma}_{ab}{\>_{}}_{n}$$+ \frac{1}{2}\<\gamma {\>_{}}_{n}$ 
  at each scale $n$, irrespective of the detailed form of 
  the spectral distance nor the gravity theory (Eq.(\ref{eq:d_N-close2})). 
  Even though it is a special case when two geometries are very close 
  to each other in ${\cal S}_N$, 
  this result would still provide us with a basic understanding about 
  what kind of geometrical  discrepancies would matter in the context of 
   the  validity of a cosmological model for the real Universe. 
  More systematic studies along this line  are certainly required, 
  which would be presented elsewhere~\cite{MS-apply}. 

  Finally, we make some remarks on  the spectral scheme in general. 
  It is a different way 
  of viewing geometrical structures  from the standard way of 
  describing  them. In this scheme, the geometrical information on a space is 
  represented by  a collection of the whole of the spectral information 
  measured by all available elliptic 
  operators on the space~\cite{MS-spectral,MS-JGRG,MS-AVE}. 
  In ordinary cosmological observations, we use a particular 
  observational apparatus so that  we naturally get only a portion of 
  the whole geometrical information on the space, 
  according  to which   apparatus (or  mathematically, which 
  elliptic operators corresponding to the apparatus) has been utilized.
  Thus, it can happen that 
   the geometry of the space cannot be  fully identified 
   by using just a single apparatus (some particular elliptic operators). 
   This consideration provides us with 
    a physical interpretation  of the isospectral manifolds. 
    From this viewpoint, there is no surprise in the existence of the 
    isospectral manifolds.
    The spectral scheme describes 
    the scale- and apparatus-dependent geometry of a space in a natural 
    manner. According to what this scheme suggests us,   
    there is no absolute  model for the real Universe, 
     rather a good model for reality depends on the observational scale 
    which we are interested in, and on the observational apparatus 
    which we rely on.  The smaller scale we pay attention to  and the more
    variety of apparatus we utilize, the closer the model Universe 
     constructed from the data approaches the real Universe.

\vskip 1cm
 
The author thanks the Ministry of Education, the Government of Japan 
 for financial support. He also thanks Inamori Foundation, Japan 
 for encouragement as well as financial support.

\makeatletter
\@addtoreset{equation}{section}
\def\theequation{\thesection\arabic{equation}}
\makeatother

\appendix

\begin{center}
\bf APPENDIX
\end{center}

\section{The zero mode}
\label{section:Appendix A}

 In our theory, the zero-mode $f_0$ of the Laplacian $\Delta$ 
 is more important than in a usual mathematical context: It 
 is directly related to the spatial volume (Eq.(\ref{eq:A-2}) below), so that 
 it is a dynamical object just as other modes.   
Hence it is essential to note   the basic facts on the zero-mode here 
for the development of our theory.  

Since $\Delta f_0=0$, it follows that $0=\int f_0 \Delta f_0 \/ 
= - \int \left(\partial f_0 \right)^2 \/$, implying that 
 $\partial_a f_0 \equiv 0$. Thus, 
\begin{equation}
\mbox{ The zero-mode of $\Delta$ is a constant function, hence there 
is no degeneracy.} 
\label{eq:A-1}
\end{equation}

Now, from Eq.(\ref{eq:normalization}) for $m=n=0$, 
it follows that $1=(f_0, f_0)=f_0^2 V$ on account of 
(\ref{eq:A-1}). Here $V$ is the $(D-1)$-volume of the space. Thus\footnote{
Here we  choose the positive squre-root. There is  no essential 
difference even when we choose the negative squre-root instead. 
For instance, Eq.(\ref{eq:<gdg>0}) in \S \ref{section:II} remains same.}
\begin{equation}
 f_0=1/\sqrt{V}\ \ .
\label{eq:A-2} 
\end{equation}
Hence  $\int f_0 \/ = \sqrt{V}$.
From Eq.(\ref{eq:normalization}) for $n=0$ with 
Eq.(\ref{eq:A-2}), we get 
\begin{equation}
\int f_m \/ = \sqrt{V} \d_{m0}\ \ .
\label{eq:A-3} 
\end{equation}
 
Eq.(\ref{eq:A-2}) can be represented as $f_0=(\int \/)^{-1/2}$. 
Taking the variation of the both sides of this equation, we easily 
get $\d f_0 = - \frac{1}{4V^{3/2}}\int g\cdot \d g \/$. With the 
help of Eq.(\ref{eq:A-2}),  we thus get 
\begin{equation}
\d f_0 = -\frac{1}{4}\<g \cdot \d g\>_{{}_0}f_0 \ \ .
\label{eq:A-4} 
\end{equation}
This formula is of  the fundamental importance to 
develop the perturbation theory 
suitable for our purpose (see {\it Appendix} \ref{section:Appendix B} 
and \S\S \ref{subsection:II-3}).

On the other hand, taking the variation of Eq.(\ref{eq:A-2}) directly, 
we obtain 
\begin{equation}
\d f_0 = -\frac{1}{2}V^{-3/2}\d V \ \ .
\label{eq:A-5} 
\end{equation}
Comparing Eqs.(\ref{eq:A-4}) and (\ref{eq:A-5}) to each other, 
we  get Eq.(\ref{eq:<gdg>0}) in \S \ref{section:II}, 
\[
\<g\cdot \d g\>_{{}_{0}}= 2 \frac{\d V}{V}\ \ .
\]

\section{Basic results of the perturbation theory}
\label{section:Appendix B} 

The perturbation theory is helpful to 
to analyze the dynamical evolution of the spectra 
$\{\l_n\}_{n=0,1,2,\cdots}$. In the present case, however, 
we need  to pay special attentions to the zero-mode as is explained in 
{\it Appendix} \ref{section:Appendix A}. 
We here derive basic formulas of the perturbation 
theory, taking care of   the zero-mode. (Up to Eq.(\ref{eq:B-7}) below, 
we mostly follow the argument in Ref.{\cite{JJS}}.)  

We consider an Hermitian operator\footnote{
 It can also be regarded as an Hermitian-operator-valued  function of $\a$.}
  $\Delta$ parameterized by $\a$ which   is analytic  in $\a$ 
 in the neighborhood of $\a =0$:
\begin{eqnarray}
\Delta &=& \Delta_0 + \a \Delta_1 + \a^2 \Delta_2 + \cdots 
                                                      \nonumber \\ 
       &=:& \Delta_0 + \a \d_\a \Delta     \ \ .             
\label{eq:B-1} 
\end{eqnarray}
The operator $\Delta$ can be arbitrary 
although we choose it  to be the Laplacian in the application.

Let $\{ (\l_n^{(\a)},\  |n {\>_{}}_\a )\}_{n=0,1,2,\cdots}$ and 
$\{ (\l_n ^{(0)},\  |n^{(0)} \> )\}_{n=0,1,2,\cdots}$ be the set of spectra and 
eigenvectors for $\Delta$ and $\Delta_0$, respectively:
\begin{eqnarray}
\Delta |n {\>_{}}_\a &=& - \l_n^{(\a)} |n {\>_{}}_\a \ \ , \nonumber \\
\Delta_0 |n^{(0)} \> &=& - \l_n^{(0)} |n^{(0)} \> \ \ .
\label{eq:B-2}
\end{eqnarray}
We understand that $\{ |n^{(0)}\> \}_{n=0,1,2, \cdots}$ is the 
normalized set of eigenvectors. On the other hand, we do not  specify the 
normalization of $\{ |n {\>_{}}_\a \}_{n=0,1,2, \cdots}$ at this stage 
(see after Eq.(\ref{eq:B-7}) below). 

We assume that both $\l_n^{(\a)}$ and $|n {\>_{}}_\a$ are analytic 
in $\a$ in the neighborhood of $\a =0$:
\begin{eqnarray}
\l_n^{(\a)} &=& \l_n^{(0)} + \a \l_n^{(1)} + \a^2 \l_n^{(2)} 
                                            + \cdots \nonumber \\
            &=:& \l_n^{(0)} + \a \d_\a \l_n \ \ ,     \nonumber \\ 
|n {\>_{}}_\a &=& |n^{(0)} \> + \a |n^{(1)} \> + \a^2 |n^{(2)} \> 
                           + \cdots    \ \ .                  
\label{eq:B-3} 
\end{eqnarray}

We introduce a projection operator $P_n$ which projects any vector to 
the sector perpendicular to $|n^{(0)}\>$:
\begin{equation}
   P_n := 1 -  |n^{(0)} \>\<n^{(0)}| 
                      = \sum_{k \neq n} | k^{(0)}\>\< k^{(0)}| \ \ .
\label{eq:B-4} 
\end{equation}

With the help of Eqs.(\ref{eq:B-1}) and (\ref{eq:B-3}), 
the first equation in (\ref{eq:B-2}) can be represented as
\begin{equation}
(\l_n^{(0)} + \Delta_0 )|n {\>_{}}_{\a} 
                  = - \a (\d_\a \l_n + \d_\a \Delta ) |n {\>_{}}_{\a}\ \ .
\label{eq:B-5} 
\end{equation}
The L.H.S. vanishes when multiplied by $\<n^{(0)}|$,  
indicating that  the R.H.S.  of Eq.(\ref{eq:B-5}) 
is perpendicular to $|n^{(0)}\>$. 
Thus, Eq.(\ref{eq:B-5}) can be represented as
\begin{equation}
(\l_n^{(0)} + \Delta_0 )|n {\>_{}}_{\a} 
           = - \a P_n (\d_\a \l_n + \d_\a \Delta ) |n {\>_{}}_{\a}\ \ 
\label{eq:B-6} 
\end{equation}
Due to the presence of $P_n$, which removes the zero-mode of the 
operator $\l_n^{(0)} + \Delta_0 $, Eq.(\ref{eq:B-6}) can be expressed as 
\begin{equation}
|n {\>_{}}_{\a} = C_n (\a)|n^{(0)}\> 
    - \frac{\a}{\l_n^{(0)} + \Delta_0}
          P_n (\d_\a \l_n + \d_\a \Delta)|n {\>_{}}_{\a} \ \ .  
\label{eq:B-7} 
\end{equation}
From here, we need to proceed in a different way from the standard 
perturbation theory in quantum mechanics. 
Noting that $\{ |n {\>_{}}_\a \}_{n=0,1,2, \cdots}$ forms an orthogonal 
set of bases due to Eq.(\ref{eq:B-2}), we can interpret  
 $C_n (\a)=\<n^{(0)} |n{\>_{}}_\a$ as the factor characterizing
  the normalization of $\{ |n {\>_{}}_\a \}_{n=0,1,2, \cdots}$. 
It should be analytic in $\a$ 
around $\a=0$ with 
$C(\a) \rightarrow 1$ as $\a \rightarrow 0$:
\begin{equation}
C_n(\a)= 1+\a c_n^{(1)} + \a^2 c_n^{(2)} + \cdots \ \ .
\label{eq:B-8} 
\end{equation}
In the standard 
perturbation theory in quantum mechanics, the inner-product of the 
states is not perturbed. Thus  any 
normalized state $|A \>$ and its perturbation $\d |A \>$ 
are always perpendicular to  each other, so that 
usually we can  set $C_n(\a)\equiv 1$~\cite{JJS}.  
In our case, on the other hand, 
the inner-product itself is perturbed through the integral-measure
 $\/$,  when we regard 
$(f_m,\ f_n)$ in Eq.(\ref{eq:normalization}) as the inner-product of 
 the ``states" $f_m$ and $f_n$. Thus we should tread $C_n(\a)$ with care 
 (see the argument after Eq.(\ref{eq:df-f})).     

Since $\<n^{(0)}|(\d_\a \l_n + \d_\a \Delta ) |n {\>_{}}_{\a}=0 $, we get
\begin{equation}
C_n(\a) \d_\a \l_n  = -\<n^{(0)}|\d_\a \Delta |n{\>_{}}_\a\ \ .
\label{eq:B-9} 
\end{equation}
Eq.(\ref{eq:B-9}) indicates that the both sides should  match as
  analytic functions around $\a=0$. Hence, taking into account 
 Eqs.(\ref{eq:B-1}), (\ref{eq:B-3}) and (\ref{eq:B-8}), we get 
\begin{eqnarray}
\l_n^{(1)}&=& -\<n^{(0)}| \Delta_1 |n^{(0)}\> \ \ , \nonumber \\
\l_n^{(2)}&=& -\<n^{(0)}| \Delta_1 |n^{(1)}\> 
                - \<n^{(0)}| \Delta_2 |n^{(0)}\> -c_n^{(1)}  \l_n^{(1)}\ \ ,
                                                       \nonumber \\
            &\cdots&                                
\label{eq:B-10}                                \\
\l_n^{(l)}&=& -\<n^{(0)}| \Delta_1 |n^{(l-1)}\> 
                  -\<n^{(0)}| \Delta_2 |n^{(l-2)}\> - \cdots 
                   -\<n^{(0)}| \Delta_l |n^{(0)}\>  \nonumber \\
             && \qquad   -c_n^{(l-1)} \l_n^{(1)} - c_n^{(l-2)} \l_n^{(2)}
                           -\cdots -c_n^{(1)}  \l_n^{(l-1)}\ \ , \nonumber \\
            &\cdots&  \ \ .  \nonumber                
\end{eqnarray}
 
Now, the R.H.S. of the second equation in (\ref{eq:B-3}) and 
the R.H.S. of Eq.(\ref{eq:B-7}) should match as vector-valued analytic 
functions around $\a=0$. Hence, taking into account  
Eq.(\ref{eq:B-8}), we get 
\begin{eqnarray}
|n^{(1)}\>&=&\sum_k |k^{(0)}\> \mu^{(1)}_{kn}\ \ , \nonumber \\
|n^{(2)}\>&=&\sum_k |k^{(0)}\> \mu^{(2)}_{kn}\ \ ,
\label{eq:B-11} 
\end{eqnarray}
 where 
\begin{eqnarray}
\mu^{(1)}_{mn}&:=&
                  \left\{
                    \begin{array}{ll}
        \frac{\< \Delta_1 {\>_{}}_{mn}}{\l_m^{(0)} - \l_n^{(0)}} \ \ 
                             & \mbox{for}\ \ m \neq n   \\
                             c_n^{(1)}  \ \ 
                             & \mbox{for}\ \ m = n\ \ . \\
       \end{array}                  
\label{eq:B-12}
        \right.                           \\ 
\mu^{(2)}_{mn}&:=&
                  \left\{
                    \begin{array}{ll}
        \frac{1}{\l_m^{(0)} - \l_n^{(0)}}
             \left(\< \Delta_2 {\>_{}}_{mn} 
                     + \l_n^{(1)} \mu^{(1)}_{mn}
                + \sum_{k} \< \Delta_1 {\>_{}}_{mk}\mu^{(1)}_{kn} 
              \right) \ \ 
                             & \mbox{for}\ \ m \neq n  \\
                             c_n^{(2)}  \ \ 
                             & \mbox{for}\ \ m = n\ \ . \\
       \end{array}                           \nonumber
        \right.                          
\end{eqnarray}

We note that Eq.(\ref{eq:B-9}) can be expressed as 
\begin{equation}
\d_\a \l_n  
= -\frac{ \<n^{(0)}|\d_\a \Delta |n{\>_{}}_\a }{ \<n^{(0)}|n{\>_{}}_\a }\ \ .
\label{eq:B-13}
\end{equation}
Thus $\d_\a \l_n$ should be independent of the normalization factor 
$C_n(\a)$.  
For example, with the help of Eq.(\ref{eq:B-11}), 
the formula for $\l_n^{(2)}$ in Eq.(\ref{eq:B-10}) turns out to be  
\begin{eqnarray}
\l_n^{(2)}  
&=& -\sum_{k \neq n} (\l_n - \l_k)\mu^{(1)}_{nk}\mu^{(1)}_{kn} 
-\< \Delta_2 {\>_{}}_{n}   \nonumber \\ 
&=& -\sum_{k \neq n}
      \frac{\< \Delta_1 {\>_{}}_{nk}\< \Delta_1 {\>_{}}_{kn} }
           {\l_k^{(0)} - \l_n^{(0)}}
-\< \Delta_2 {\>_{}}_{n}   \ \ ,
\label{eq:B-14}
\end{eqnarray}
which is independent of $c_n^{(l)}$ ($l=1,2,\cdots $).

Mostly  $\a$ is identified with a time-function $t$ in this paper. 
When we interpret   the variation $\d Q$ of any quantity $Q$ 
 to be  a derivative of a corresponding  
function $Q(\a)$  at $\a=0$, only the terms of order $O(\a)$ 
(quantities with superscript indices 
$(0)$ and  $(1)$) in the above formulas are important.


\end{document}